\documentclass[namedreferences]{solarphysics}

\usepackage[hyperref,optionalrh,showbiblabels]{spr-sola-addons} % For Solar Physics 
\usepackage{graphicx}        % For eps figures, newer & more powerfull
\usepackage{color}           % For color text: \color command
\usepackage{breakurl}        % For breaking URLs easily trough lines
\usepackage{multicol}

\newcommand{\aap}{    {\it Astron. Astrophys.}}

\newcommand{\apj}{    {\it Astrophys. J.}}

\newcommand{\apss}{   {\it Astrophys. Space Sci.}}

\newcommand{\grl}{    {\it Geophys. Res. Lett.}}

\newcommand{\jgr}{    {\it J. Geophys. Res.}}

\newcommand{\solphys}{{\it Solar Phys.}}
 
\newcommand{\ssr}{    {\it Space Sci. Rev.}} 

\newcommand{\araa}{ {\it Annual Review of Astronomy and Astrophysics}}

\chardef\us=`\_

\begin{document}

\begin{article}
\begin{opening}

\title{DH Type II Radio Bursts During Solar Cycles 23 and 24: Frequency-dependent Classification and their Flare-CME Associations}
 %{\it Solar Physics}
\author[addressref={aff1,aff2},corref,email={binalp@prl.res.in}]{\inits{B. D.}\fnm{Binal D.}~\lnm{Patel}\orcid{https://orcid.org/0000-0001-5582-1170}}%\sep
\author[addressref=aff1]{\inits{B.J.}\fnm{Bhuwan}~\lnm{Joshi}\orcid{https://orcid.org/0000-0001-5042-2170}}%\sep

\author[addressref={aff3,aff4}]{\inits{K.-S.}\fnm{Kyung-Suk}~\lnm{Cho}\orcid{https://orcid.org/0000-0003-2161-9606}}%\sep

\author[addressref={aff3,aff4}]{\inits{R.-k.}\fnm{Rok-Soon}~\lnm{Kim}\orcid{https://orcid.org/0000-0002-9012-399X}}

\address[id=aff1]{Udaipur Solar Observatory, Physical Research Laboratory, Udaipur 313001, India}
\address[id=aff2]{Indian Institute of Technology, Gandhinagar 382355, India}

\address[id=aff3]{Space Science Division, Korea Astronomy and Space Science Institute, Daejeon 34055, Republic of Korea}
\address[id=aff4]{Department of Astronomy and Space Science, University of Science and Technology, Daejeon 34113, Republic of Korea}

\runningauthor{B. D. Patel \textit{et al.}}
\runningtitle{DH-type II bursts during Solar Cycles 23 and 24}

\begin{abstract}
We present the characteristics of DH type II bursts for the Solar Cycles 23 and 24. The bursts are classified according to their end frequencies into three categories, i.e. Low Frequency Group (LFG; 20~kHz $\leq$~{\it f}~$\leq$ 200 kHz), Medium Frequency Group (MFG; 200 kHz $<$~{\it f}~$\leq$ 1 MHz), and High Frequency Group (HFG; 1 MHz $<$~$f$~$\le$ 16 MHz). We find that the sources for LFG, MFG, and HFG events are homogeneously distributed over the active region belt. Our analysis shows a drastic reduction of the DH type II events during Solar Cycle 24 which includes only 35\% of the total events (i.e. 179 out of 514). Despite having smaller number of DH type II events in the Solar Cycle 24, it contains a significantly higher fraction of LFG events compared to the previous cycle (32\% \textit{versus} 24\%). However, within the LFG group the cycle 23 exhibits significant dominance of type II bursts that extend below 50 kHz, suggesting rich population of powerful CMEs travelling beyond half of the Sun-Earth distance. The events of LFG group display strongest association with faster and wider (more than 82\% events are halo) CMEs while at the source location they predominantly trigger large M/X class flares (in more than 83\% cases). Our analysis also indicates that CME initial speed or flare energetics are partly related with the duration of type II burst and that survival of CME associated shock is determined by multiple factors/parameters related to CMEs, flares, and state of coronal and interplanetary medium. The profiles relating CME heights with respect to the end frequencies of DH type II bursts suggest that for HFG and MFG categories, the location for majority of CMEs ($\approx$65\%--70\%) is in well compliance with ten-fold Leblanc coronal density model, while for LFG events a lower value of density multiplier ($\approx$3) seems to be compatible. 
\end{abstract}

\end{opening}

\section{Introduction}
 
Coronal mass ejections (CMEs) along with associated eruptive flares are among the most spectacular, large-scale phenomena on the Sun and in the solar system. A flare is essentially characterized by the catastrophic and explosive energy release of the order of $10^{27}$--$10^{32}$ ergs within tens of minutes in localized regions of the solar atmosphere. After initial activation and subsequent ejection of the magnetized plasma from the source region during an eruptive flare, the large-scale structure of CME is seen to propagate outward through the higher coronal layers which eventually enters into the interplanetary medium. CMEs are known to be associated with variety of other important phenomena, such as, solar energetic particle (SEP) event, interplanetary (IP) shock, geo-magnetic storm (GS), \textit {etc.}, which form important ingredients of the contemporary space weather research.

The powerful CME events are known to generate shocks in the coronal and interplanetary medium which are identified as type II bursts in radio observations \citep[e.g.,][]{2001JGR...10629219G, 2007ApJ...663.1369R, 2018SoPh..293..107J}. A type II burst is classically identified in dynamic spectra as slowly drifting feature toward lower frequencies \citep{1963ARA&A...1..291W}. The type II emission is essentially an escaping electromagnetic radiation in the radio domain; hence they arrive to the observer at Earth in $\approx$8.3 minutes, providing the advance warning of arrival of interplanetary shock. The onset frequency of a type II reflects the heliocentric distance at which the underlying shock is formed \citep[e.g.,][]{2007SoPh..244..167P}. Various observations and theoretical models reveal that the type II emissions are generated in the upstream region of CME driven shock \citep[see][]{2003JGRA..108.1126K}. It has also been suggested that, in the near-Sun region, CME nose and CME-streamer interaction can be distinct sites of multiple type II bursts \citep[see][]{2011A&A...530A..16C}. 

Depending on the energetics of CMEs, the type II bursts are observed in metric (m; 30 MHz $\leq$~{\it f} $\leq$ 300 MHz), decametric-hectometric (DH; 300 kHz $\leq$~{\it f} $\leq$ 30 MHz) and kilometric (km; 30 kHz $\leq$~{\it f} $\leq$ 300 kHz) wavelength domains. The different variants of type II bursts have been schematically illustrated and explained in \cite{2006GMS...165..207G, 2010nspm.conf..108G, 2011pre7.conf..325G}. The extension or origin of a type II radio bursts in the DH domain implies the cases of stronger MHD shocks propagating from the inner corona and entering the IP medium \citep[e.g.,][]{2007ApJ...663.1369R, 2018Ap&SS.363..126S}. Hence, the study of shocks in DH domain together with their associated CME-flare events becomes extremely important to infer not only the propagation characteristics of CMEs but to develop their forecasting tools; this study is an important attempt to broaden our understanding on the above aspects through a comprehensive statistical analysis.  

The present investigation becomes feasible due to availability of long-term, almost uninterrupted data set of IP type II radio bursts, covering almost three previous cycles (1994 to date); thanks to Radio and Plasma Wave Experiment \citep[WAVES,][]{1995SSRv...71..231B} on-board Wind spacecraft. Importantly, WAVES provides unique opportunity to observe DH type II bursts in the wavelength domain of 20 kHz--13.825 MHz, which cannot be observed from the ground due to ionospheric cut-off at $\lessapprox$15 MHz. These low radio frequency observations extending much below kilometric range (i.e. $\lessapprox$300 KHz) are of immediate applicability in tracking the locations of energetic CMEs \citep{2005JGRA..11012S07G}. Contextually, the metric type II radio bursts that extend to DH range and have origin in the western hemisphere of the Sun show strong association with SEP events \citep[e.g.,][]{2004ApJ...605..902C, 2018arXiv181011173G}. It is also noteworthy that the end frequency of DH type II bursts is statistically related to the geo-effectiveness of CMEs: the lower the type II burst end frequency, the higher the possibility of having a stronger
storm \citep[e.g.,][]{2019SoPh..294...47S}. The above results also point toward the usefulness of simple DH type II parameters, \textit{viz.} end frequency and duration, in examining the IP propagation, near-Earth consequences, and terrestrial response of CMEs.

In this paper, we present statistical study of DH type II bursts occurred during solar cycles 23 and 24. The novelty of this analysis lies in classifying DH type II bursts in terms of their end frequencies. As such, the end frequencies of these IP bursts provide a quantitative estimation of the distance up to which a shock can survive and hence have implications in understanding the energetics and propagation characteristics of CMEs. In section~\ref{sec:data_and_char}, we present a brief description of the data sources. This section gives an account of the classification scheme for DH type II bursts along with relevant physical interpretation. The section~\ref{char_LMH} presents the characteristics of the bursts for each frequency group and their association with properties of CMEs and flares. We also compare the height of CME leading edge estimated from the type II observations with direct coronographic measurements and comment on the applicability of CME propagation models.

\section{Observational Data and Basic Characteristics of DH Type II Radio Bursts}
\label{sec:data_and_char}
\subsection{Data Sources}
For the present analysis, we have obtained data from the following sources.
\begin{enumerate}

\item Wind/WAVES Type-II Burst Catalogue\footnote{See \url{http://cdaw.gsfc.nasa.gov/CME\_list/radio/waves\_type2.html.}}:

This catalogue contains the list of DH type II bursts identified by Radio and Plasma Wave Experiment \citep[WAVE,][]{1995SSRv...71..231B} on board Wind spacecraft (1994--present) and SWAVE \citep{2008SSRv..136..487B} on board Solar TErrestrial RElations Observatory (STEREO; 2006--present). Wind/WAVES is the primary instrument providing continuous observations of low-frequency radio emission from the Sun since its launch in the frequency range of 14 MHz--20 kHz. The low frequency receiver (RAD--1) observes in the frequency range 20 kHz to 1.04 MHz while the high frequency receiver (RAD--2) covers the frequency range from 1.075 to 13.825 MHz. SWAVE\footnote{See \url{https://swaves.gsfc.nasa.gov/swaves_instr.html.}} has further complemented Wind/WAVES measurements by extending the frequency range upto 16~MHz. However, it should be noted that SWAVE instruments (on board STEREO A and B satellites) are located at different vantage points for Sun's observations (one ahead of the Earth and other behind it), the fine structure and intensity of a type II may vary between observations from A and B spacecraft. The data from SWAVE is available from November 11, 2006. This analysis includes 68 DH type II events identified by SWAVE during the period 2006-2017. The combined observations from the two instruments are contained in the Wind/WAVES type II burst catalogue which is generated and maintained at the Coordinated Data Analysis Workshop (CDAW) data center by NASA and The Catholic University of America \citep{2019SunGe..14..111G}. The catalogue also identifies CMEs associated with DH type II events with additional information on the source region parameters, such as, heliographic coordinates, active region number, and GOES class of associated flare. Wind/WAVE catalogue is updated up to September 2017.
	
\item Solar and Heliospheric Observatory (SOHO) Large Angle and Spectroscopic Coronagraph (LASCO) CME Catalogue\footnote{See \url{http://cdaw.gsfc.nasa.gov/CME\_list/.}}: 
We explore the properties of CMEs at the near-Sun region associated with DH type II radio bursts on the basis of parameters derived from SOHO/LASCO observations and published in this catalogue \citep{2009EM&P..104..295G}. The LASCO images the white light corona from 2 R$_\odot$--30 R$_\odot$. The catalog is generated and maintained at the Coordinated Data Analysis Workshop (CDAW) Data Center by NASA and The Catholic University of America in cooperation with the Naval Research Laboratory. From this catalogue, we obtained relevant physical parameters that describe kinematic properties of CME, such as, linear speed, angular width, and height-time data. 
\end{enumerate}		

\begin{figure}
\centering
\includegraphics[width=\textwidth]{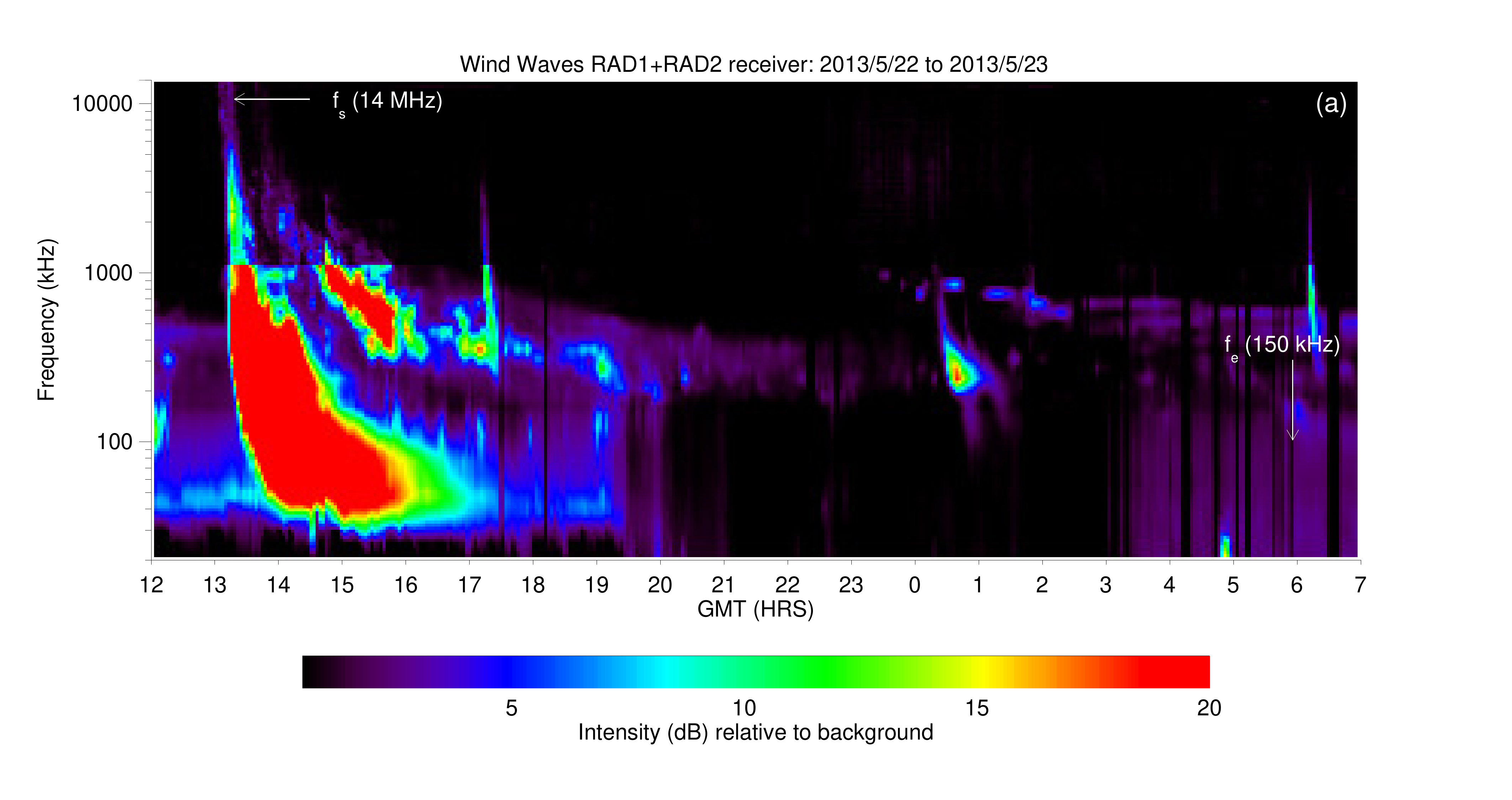}
\includegraphics[width=\textwidth]{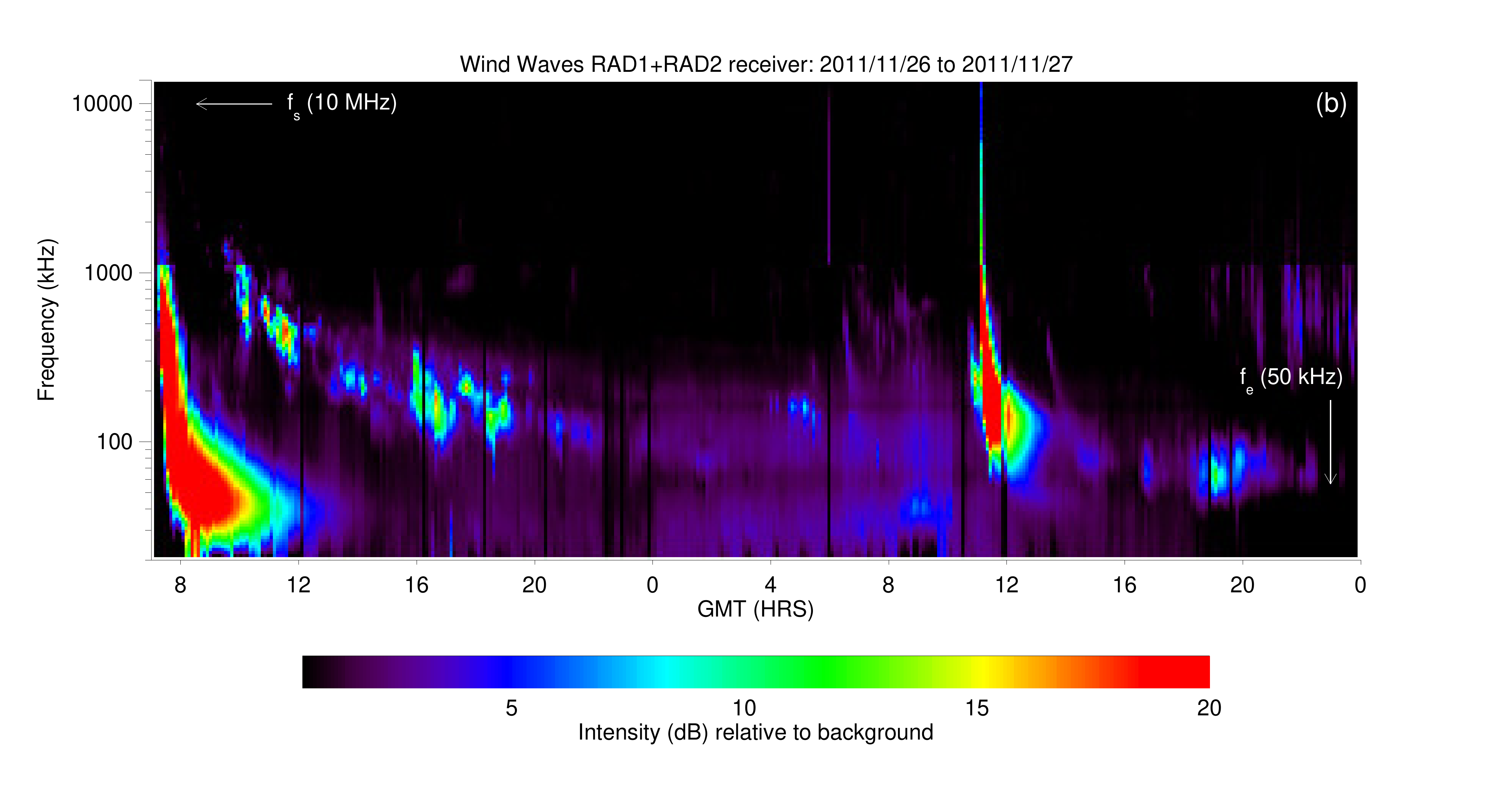}
\caption{Dynamic radio spectrum obtained from WAVES experiment on-board Wind spacecraft, which operates in the frequency range of 14 MHz to 20 kHz. A typical example of DH type II radio burst associated with CME on 2013 May 22 (panel a) and 2011 November 26 (panel b). Start and end frequencies of the radio burst are annotated by f$_{s}$ and f$_{e}$, respectively.}
\label{fig:spectrum}
\end{figure}

\begin{figure}
\centering
\includegraphics[width=\textwidth]{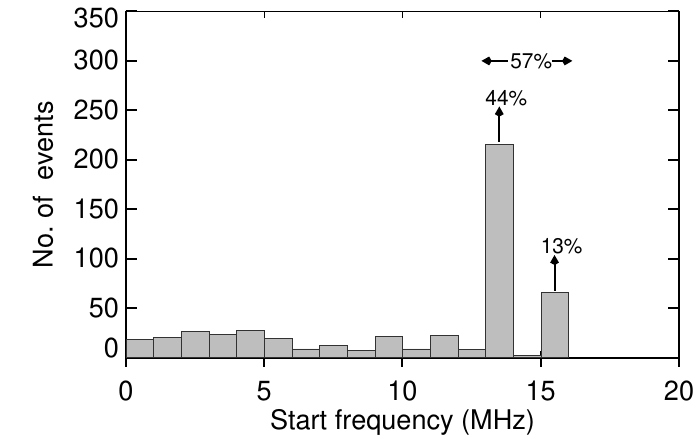}
\caption{Distribution of DH type II radio bursts in terms of their start frequencies for all the events combined for Solar Cycles 23 and 24.}
\label{fig:hist_start_frq}
\end{figure}

Based on WIND/WAVE observations, our primary data set constitutes a total of 514 DH type II bursts that occurred during January 1997--September 2017.
In Figure~\ref{fig:spectrum}, we present two examples of DH type II radio burst observed by Wind/WAVES. In this figure, we mark the start and end frequencies by arrows and denote them by $f_{\rm s}$ and $f_{\rm e}$ respectively. The interval between start and end timings of the burst defines its duration ($\tau_{\rm DH}$ = $t_{\rm e}$--$t_{\rm s}$). Also, noteworthy is the fact that the DH type II shown in Figure~\ref{fig:spectrum}a has starting frequency of 14 MHz which happens to be the observing limit of the WAVE instrument. In several occasions, this also implies that the observed type II is simply the extension of a burst starting at frequencies above 14 MHz. In order to have a clear understanding of the observed DH type II events in terms of the frequency at which the bursts originated or first detected by the concerned instruments (WAVES or SWAVES), we present a histogram of the starting frequencies in Figure~\ref{fig:hist_start_frq}. The histogram reveals that the bars corresponding to the limiting frequency range of Wind/WAVES (i.e. 13--14 MHz) and SWAVES (i.e. 15--16 MHz) represent 44\% and 13\% events, respectively. 
From this statistics, it is clear that a sizable population of DH type II events are either starting near 14--16~MHz or appearing as an extension of type II emission starting at higher frequencies.

\subsection{Origin of DH Type II Radio Bursts and Classification Scheme}
\label{sec:origin-classification}

\begin{figure}
	\includegraphics[width=\textwidth]{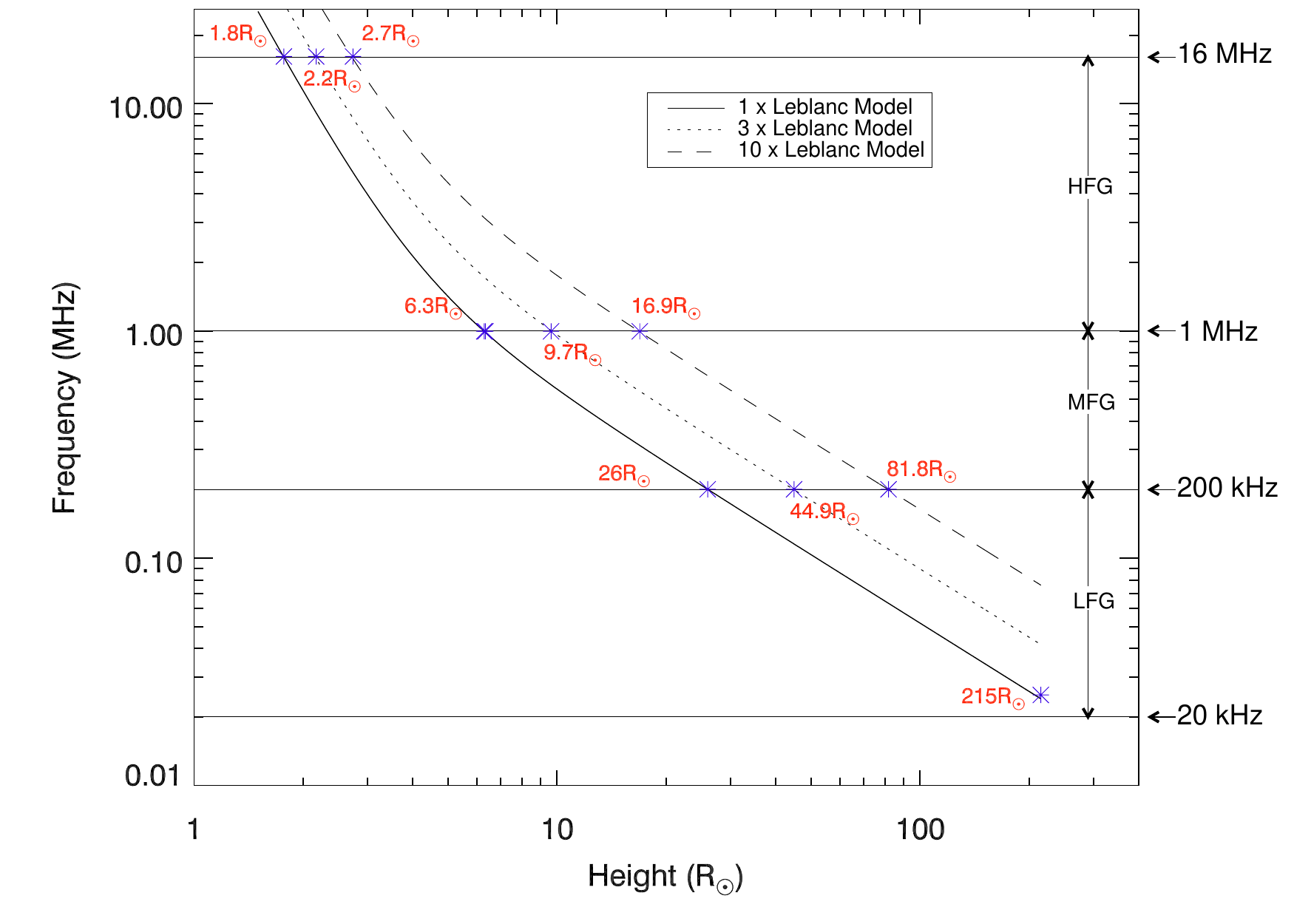}
	\caption{Variation of heliocentric burst height (R$_\odot$) with plasma frequency estimated from Leblanc electron density distribution model. Solid, dotted and dashed lines represent one-fold, three-fold, and ten-fold models, respectively. We consider three groups of DH type II radio bursts, depending upon their end frequency in the range of 1--16~MHz, 200~kHz--1~MHz, and 20~kHz--200~kHz which are denoted as High Frequency Group (HFG), Medium Frequency Group (MFG), and Low Frequency Group (LFG), respectively.}
\label{fig:labnalc}
\end{figure}

It is generally accepted that type II bursts are created by a propagating shock \citep{1985srph.book..333N, 1987JGR....92.9869C, 1999JGR...10416979R, 2001JGR...10629989R}. The electron beams produced in the shock excite Langmuir waves which subsequently get converted into escaping radio waves \citep{1995LNP...444..183M}. The emission mechanism is plasma emission near the fundamental and second harmonic frequencies. A typical feature of type II burst is slow frequency drift toward lower frequencies, implying fall in the electron density. This aspect is attributed to the burst driver moving in the solar atmosphere and, thus, toward larger heliocentric distances. A frequency-height relationship, inferred from the atmospheric density model by \cite{1998SoPh..183..165L}, is presented in Figure~\ref{fig:labnalc}. It is important to note that the height estimations from the atmospheric density model provides a ``coarse'' estimation only as the atmospheric density through which shock is propagating depends strongly on coronal and interplanetary conditions. Therefore, to fit the observational data, we consider a multiplier (or suitable ``fold'') to the basic atmospheric density model. In Figure~\ref{fig:labnalc}, the solid, dotted and dashed lines illustrate one--, three--, and ten--fold Leblanc model. The frequency-height relationship suggests that the observing window of 16 MHz--20 kHz essentially represents a large heliocentric distance from $\approx$2 R$_\odot$ to 1 AU. Therefore, in order to explore the characteristic of DH type II radio bursts, we divide this vast frequency range into three groups: Low Frequency Group (LFG; 20~kHz $\leq$~{\it f}~$\leq$ 200 kHz), Medium Frequency Group (MFG; 200 kHz $<$~{\it f}~$\leq$ 1 MHz), and High Frequency Group (HFG; 1 MHz $<$~$f$~$\le$ 16 MHz). Since end frequency of a type II burst provides a quantitative estimation on the distance up to which a shock can survive, we have classified the events into three groups based on their end frequencies. 

Figure~\ref{fig:labnalc} further suggest that HFG and MFG events essentially suggest shocks terminating within the lower and upper coronal heights (9.7 R$_\odot$ and 44.9 R$_\odot$, as per three-fold Leblanc model), respectively. On the other hand, the events under LFG group represent shock travelling in the interplanetary medium.

\begin{figure}
\centering
\includegraphics[width=1.0\textwidth]{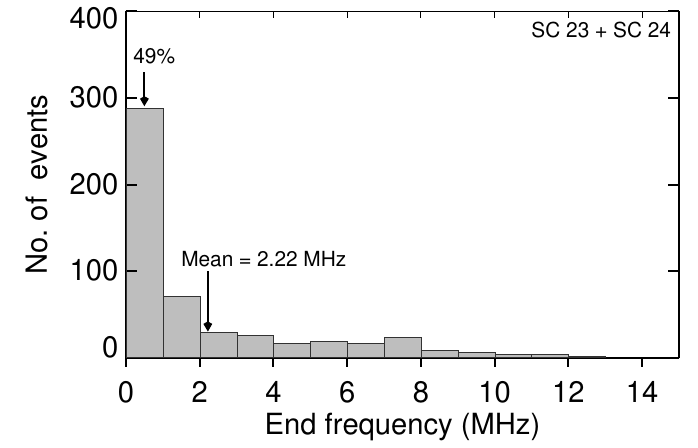}
\caption{Distribution of DH type II radio bursts in terms of their end frequencies for all the events combined for Solar Cycles 23 and 24.}
\label{fig:hist_end_frq_all}
\end{figure}

\begin{figure}
\centering
\includegraphics[width=1.0\textwidth]{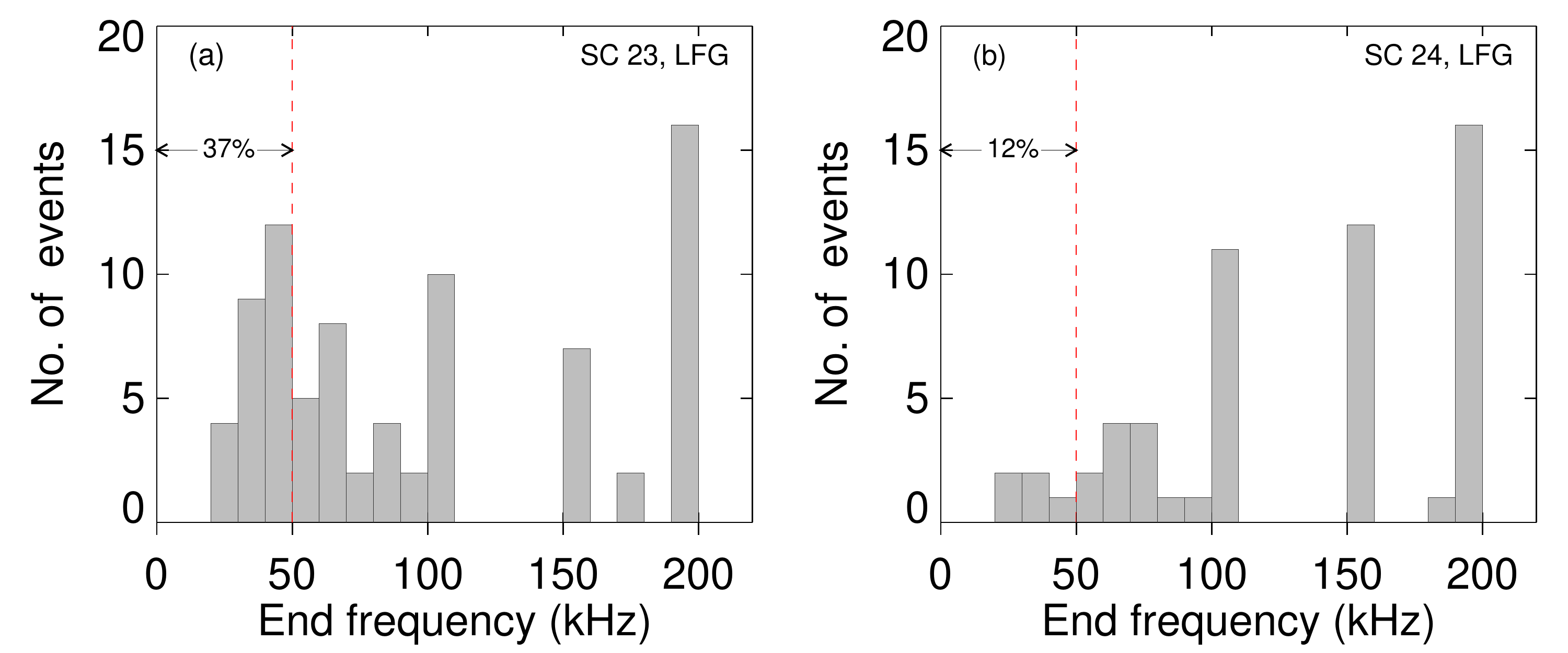}
\caption{Histogram showing the distribution of end frequencies for events of LFG categories for Solar Cycles 23 (panel a) and 24 (panel b). The red line in both the panel, represents the end frequency at 50 kHz.}
\label{fig:hist_end_frq_lfg}
\end{figure}

In Figure~\ref{fig:hist_end_frq_all}, we present histogram of ending frequency for  the DH type II events. The histogram reveals that 49\% of events fall below 1 MHz frequency bin. Notably, this lowest frequency bin also include LFG events which represents shock travelling beyond $\approx$45 R$_\odot$ and, therefore, are of great importance for space-weather prospective. However, the combined ending frequency distribution in Figure~\ref{fig:hist_end_frq_all} does not provide a detailed account for the LFG group. Therefore, we present separate histogram exclusively for LFG group in Figure~\ref{fig:hist_end_frq_lfg} separately for Solar Cycle 23 (panel a) and 24 (panel b). Here we further examine the event counts into two subgroups: 200--50 kHz and 20--50 kHz as a frequency of $\approx$50 kHz (see vertical dashed line in Figure~\ref{fig:hist_end_frq_lfg}) roughly corresponds to heliocentric distance of 0.5~AU (see Figure~\ref{fig:labnalc}). This suggests that 37\% and 12\% of the LFG events were able to produce the shock beyond 0.5 AU for Solar Cycles 23 and 24, respectively.

\subsection{Overview of DH Type II Activity During Solar Cycles 23 and 24}
\begin{figure}[h!]
   \includegraphics[width=\textwidth]{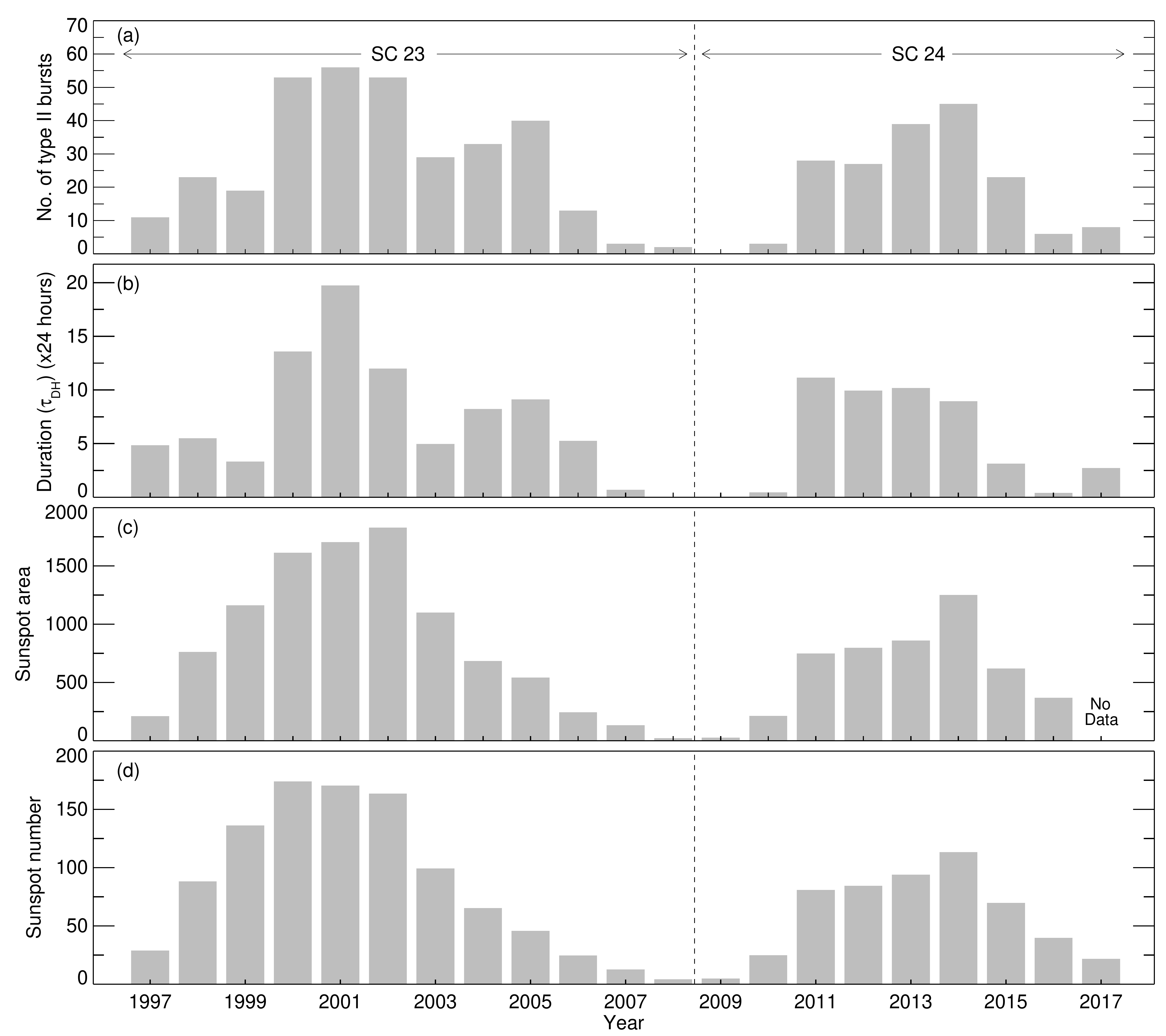}
   \caption{Histogram showing distribution of DH type II events (panel a), yearly cumulative duration ($\tau_{\rm DH}$) of DH type II events (panel b), yearly sunspot area measured in units of millionth of a hemisphere (panel c), and yearly sunspot number (panel d).}
   \label{fig:histogram_event_dur}
\end{figure}

\begin{figure}
   \includegraphics[width=\textwidth]{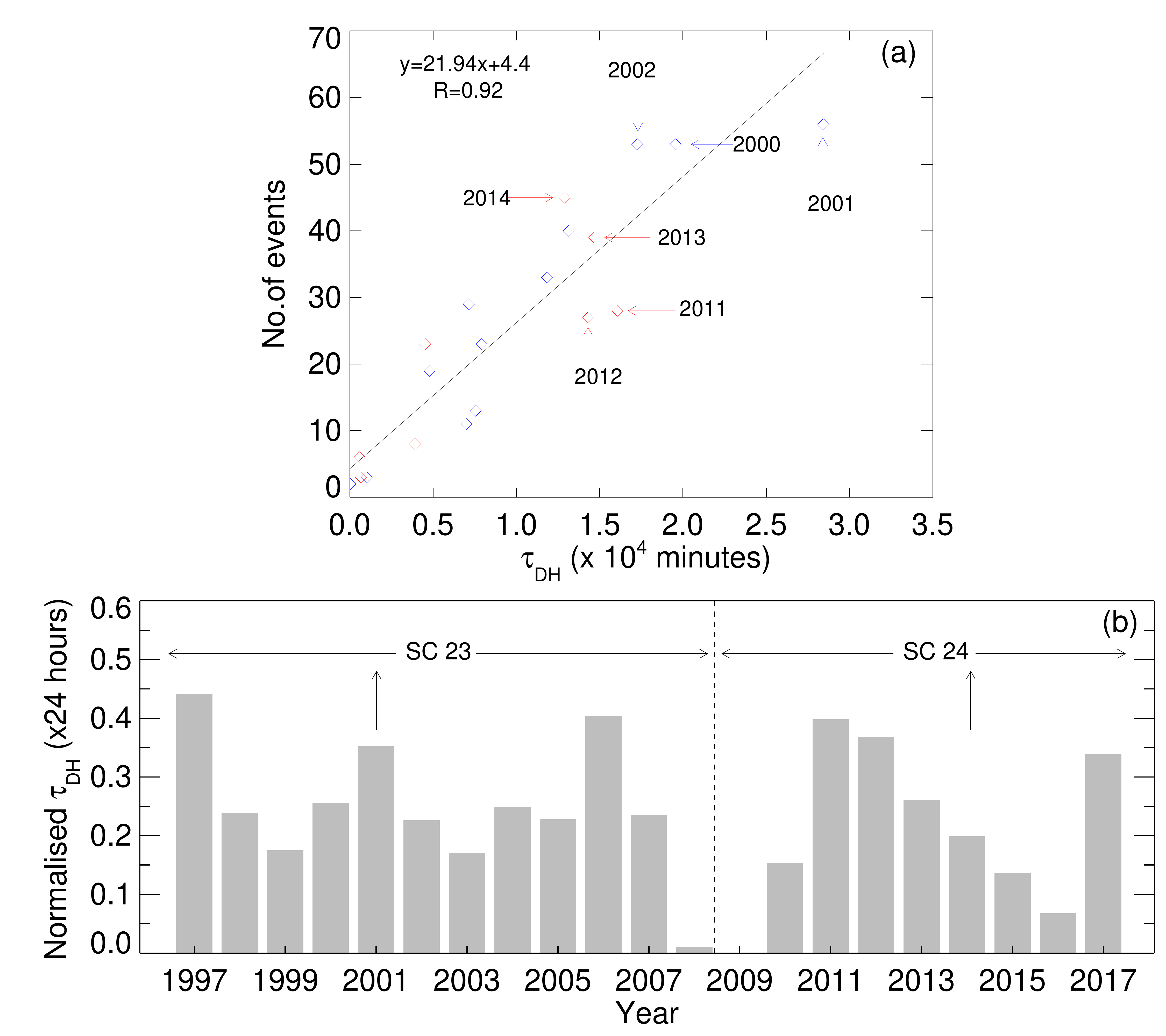}
   \caption{Correlation of yearly DH type II occurrence and yearly duration of DH type II ($\tau_{\rm DH}$) bursts with blue and red symbols representing data for Solar Cycles 23 and 24, respectively (panel a). Histogram showing distribution of normalised duration (i.e. ratio of yearly event duration and event counts) of DH type II events (panel b). Upward vertical arrows represents maximum activity year for solar cycle.}
   \label{fig:corr_event_dur}
\end{figure}

\begin{figure}
   \includegraphics[width=\textwidth]{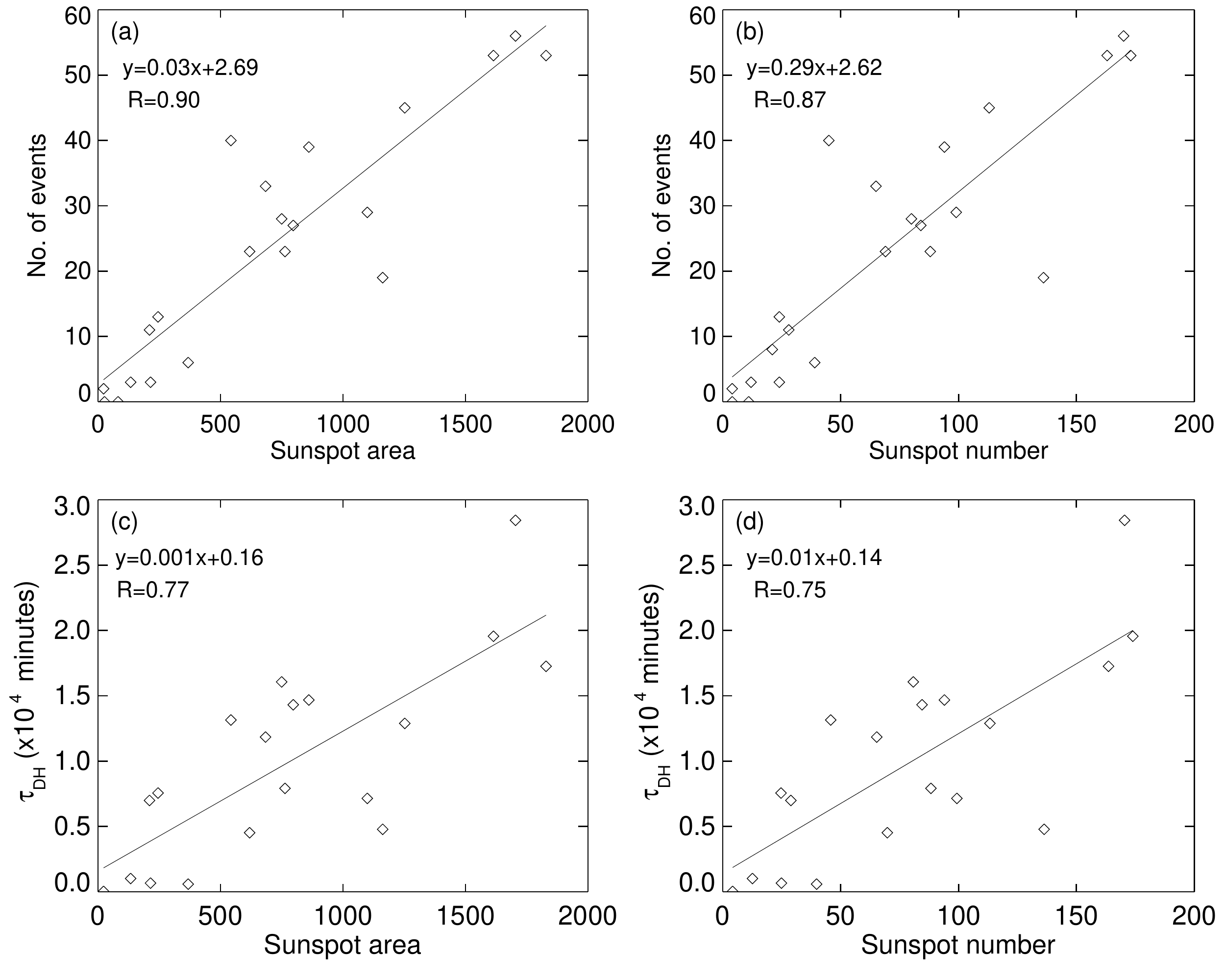}
   \caption{Correlation of number of DH type II of events with sunspot area (panel a), and sunspot number (panel b); Correlation of yearly  duration of DH type II events with  yearly sunspot area (panel c), and yearly sunspot number (panel d). }
   \label{fig:corr_event_ssn_sa}
\end{figure}

In Figure~\ref{fig:histogram_event_dur}, we present histogram showing annual occurrence of DH type II radio bursts from 1996 to 2017, covering Solar Cycles 23 and 24. We represent the activity in two ways: the event numbers (Figure~\ref{fig:histogram_event_dur}a) and total duration of events (Figure~\ref{fig:histogram_event_dur}b). As expected, the two parameters (event number and duration) correlate very well with a correlation coefficient r=0.92 (Figure~\ref{fig:corr_event_dur}a). In Figure~\ref{fig:histogram_event_dur}c and d, we present histograms of yearly sunspot number\footnote{Source: WDC-SILSO, Royal Observatory of Belgium, Brussels.} and area\footnote{Source: Royal Observatory, Greenwich--USAF/NOAA Sunspot Data.}. The sunspot area has been measured in units of millionths of a hemisphere. We find an excellent correlation between DH type II parameters and sunspot indices (Figure~\ref{fig:corr_event_ssn_sa}). In all the cases, sunspot area shows higher correlation with DH type II parameters over sunspot number.

The representation of type II burst activity in terms of event duration reveals a few peculiarities which requires further attention. To clarify this, we plot histogram of yearly normalised durations (i.e. ratio of yearly event duration and event counts plotted in Figure~\ref{fig:histogram_event_dur}a and b, respectively) in Figure~\ref{fig:corr_event_dur}b. It is interesting to note that during Solar Cycle 24, there is a systematic decrease in number of events of longer durations from rise to maximum phase of solar cycle i.e. 2011--2014 (corresponding years are also annotated in Figure~\ref{fig:corr_event_dur}a). Notably, the maximum activity year 2014 significantly lacks events of longer duration. Also, during the high activity phase of Solar Cycle 23 (2000--2002), the normalised duration of type II bursts observed in the years 2000 and 2002 is significantly lower in comparison to the maximum activity year of 2001 (Figure~\ref{fig:corr_event_dur}b); this difference is not significant when considering yearly number of type II events (Figure~\ref{fig:histogram_event_dur}a). This also suggests that the histogram bars of DH type II events corresponding to the years 2000 and 2002 consist of events with shorter duration (corresponding years are annotated in Figure~\ref{fig:corr_event_dur}a).

\section{DH Type II Characteristics for High, Medium and Low Frequency Domains}
\label{char_LMH}
As discussed in Section~\ref{sec:origin-classification}, to explore the characteristics of DH type II radio bursts, we classify the events in three frequency domains: LFG (20~kHz $\leq$~{\it f}~$\leq$ 200 kHz), MFG (200 kHz $<$~{\it f}~$\leq$ 1 MHz), and HFG (1 MHz $<$~$f$~$\le$ 16 MHz). 
In Table~\ref{table:occ_23_24}, we provide number of events for each group. We note a drastic reduction of DH type II events during Solar Cycle 24 which includes only $\approx$35\% of the total events. In the following subsections, we discuss the properties of DH type II and associated phenomena in detail.

\begin{table}
	\caption{Occurrence of DH type II events during Solar Cycles 23 and 24 for all the events.}
	\begin{tabular}{ccccc}

	\hline
		Solar Cycle & 	\multicolumn{4}{c} {Number of events} \\
      \cline{2-5}

          & LFG  & MFG & HFG & Total\\
        &\textit{f} $ \leq$ 200 kHz & 200 kHz $<$ \textit{f} $\leq$ 1 MHz & 1 MHz $<$ \textit{f} $\leq$16 MHz&\\
          \hline
           23 & 81(24\%) & 96 (29\%)&158 (47\%) &335 (65\%)\\
           24 & 57 (32\%) & 54 (30\%) & 68 (38\%) &179 (35\%)\\
           23 +  24 &  138 (27\%)& 150 (29\%) & 226 (44\%)&514\\
          \hline

\end{tabular}
\label{table:occ_23_24}
\end{table}

\subsection{Occurrence and Source Location}
\begin{figure}
\centering
\includegraphics[width=\textwidth]{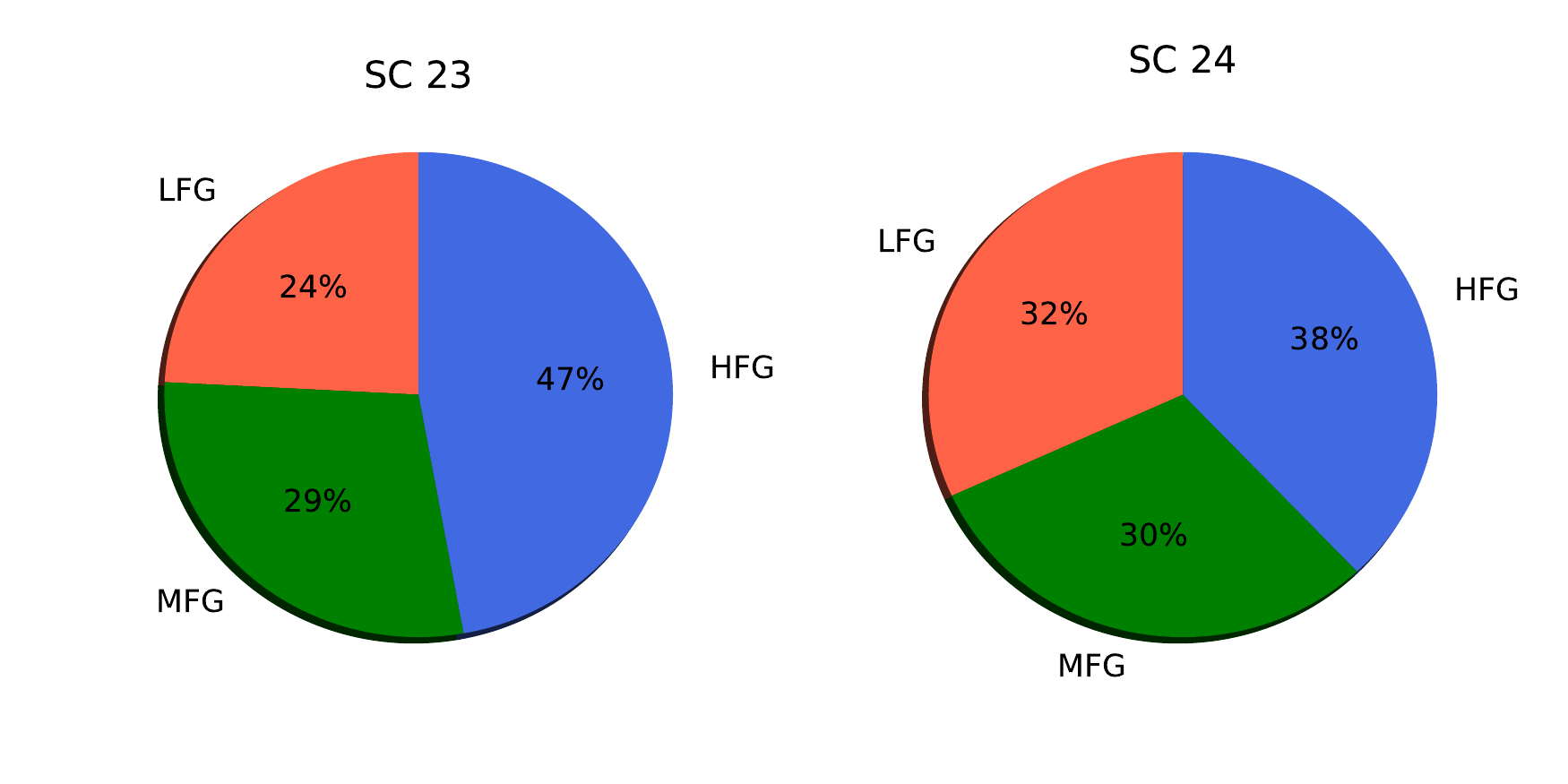}
\vspace{-0.5cm}
\caption{Pie chart showing a comparison between number of DH type II radio bursts under the LFG (red), MFG (blue) and HFG (green) during Solar Cycle 23 (left) and 24 (right). The end frequency ($f$) varies in the range of 20~kHz $\leq$~\textit{f}~$\leq$ 200 kHz for LFG and 200 kHz $<$~\textit{f}~$\leq$ 1 MHz for MFG and 1 MHz $<$~~\textit{f}~$\le$ 16 MHz for HFG.}
\label{fig:pi_LFG_HFG}
\end{figure}

In Figure~\ref{fig:pi_LFG_HFG}, we present pie charts showing the occurrence of DH type II events for LFG, MFG, and HFG categories, separately for Solar Cycle 23 (left panel) and 24 (right panel). We find that, for both the cycles, the HFG category constitutes the highest number of events. It is important to note that although Solar Cycle 24 is significantly poorer than the previous cycle in terms of overall DH type II activity (179 \textit{versus} 335; see Table~\ref{table:occ_23_24}), it contains significantly higher fraction of events under LFG category. As shown in Figure~\ref{fig:pi_LFG_HFG}, the portions of LFG events are 32\% and 24\% for Solar Cycle 24 and 23, respectively. This interesting result has important implications: in spite of smaller number of DH type II events, cycle 24 is rich in terms of producing CMEs that are able to derive shocks up to larger heliocentric distances in comparison to cycle 23.

\begin{figure}
	\includegraphics[width=11cm]{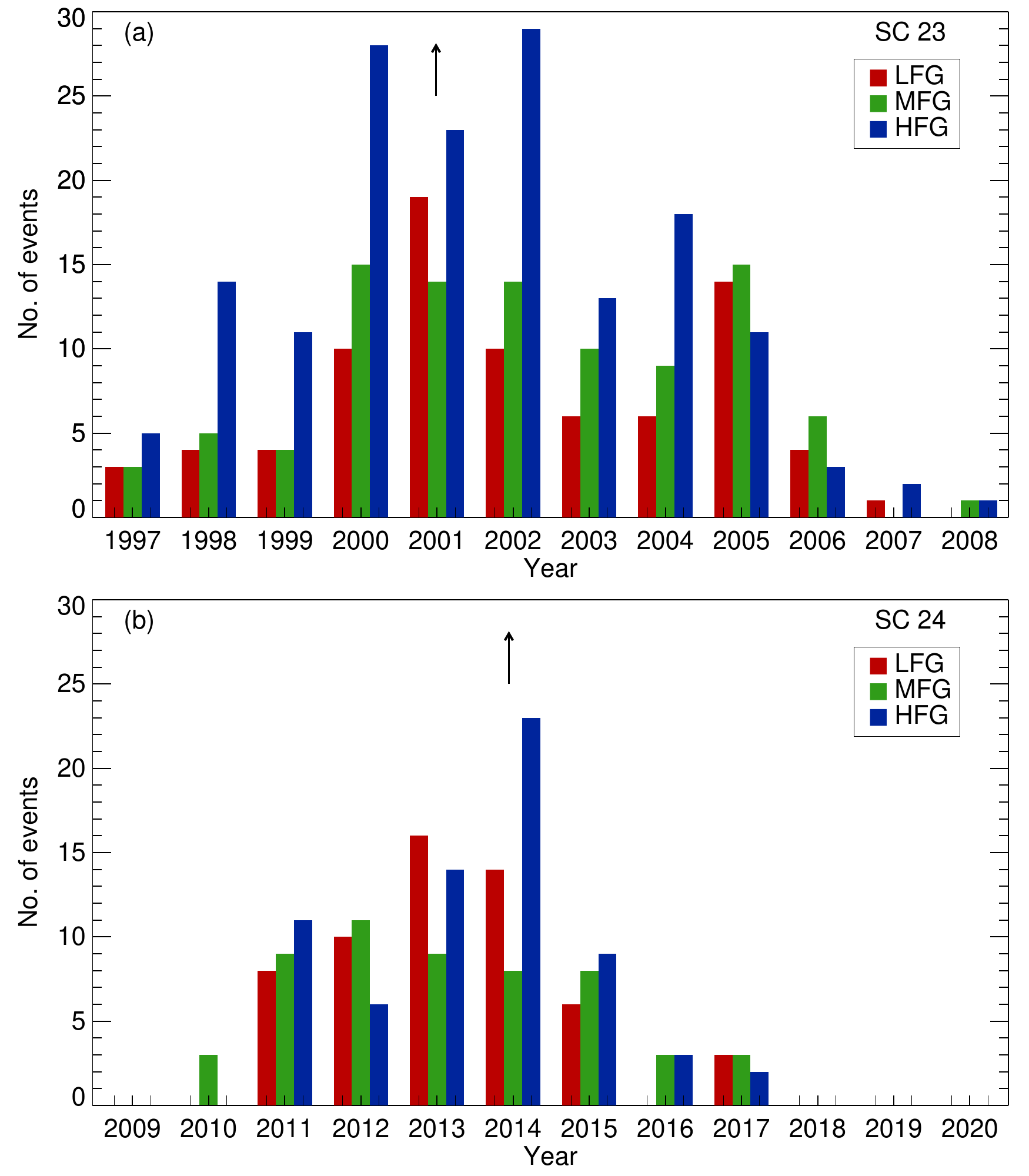}
	\caption{Histogram showing the annual occurrence of DH type II radio bursts during Solar Cycle 23 (panel a) and 24 (panel b). The arrow represent the maximum year during Solar Cycle 23 and 24. The red, green, and blue bar represent the LFG, MFG, and HFG group, respectively. }
	\label{fig:hist_DH_typeII}
\end{figure}

In Figure~\ref{fig:hist_DH_typeII}, we present year-wise histogram of DH type II bursts for LFG (red), MFG (green), and HFG (blue) categories for Solar Cycle 23 (Figure~\ref{fig:hist_DH_typeII}a) and 24 (Figure~\ref{fig:hist_DH_typeII}b). These histograms suggest some interesting features which point toward different behaviour of the two cycles. During most of the years of Solar Cycle 23 (spanning 8 years in continuity from 1997 to 2004), the events under HFG category exceed that of MFG and LFG categories. Only during the years 2005 and 2006, which falls in the minimum phase of the cycle, we find the dominance of LFG and MFG categories over the HFG. On the other hand, Solar Cycle 24 shows dominance of LFG events over the HFG ones even during the rise and high activity phases of the cycle (see histogram bars corresponding to the years 2012 and 2013). A significant difference is noticeable in the DH type II activity of the three categories only during the year of solar cycle maximum in 2014. Further, we note that during the early rise and decline phases of the cycle, the difference of activity level for the three categories is not very prominent (i.e. years 2011 and 2015).

\begin{figure}
\includegraphics[width=\textwidth]{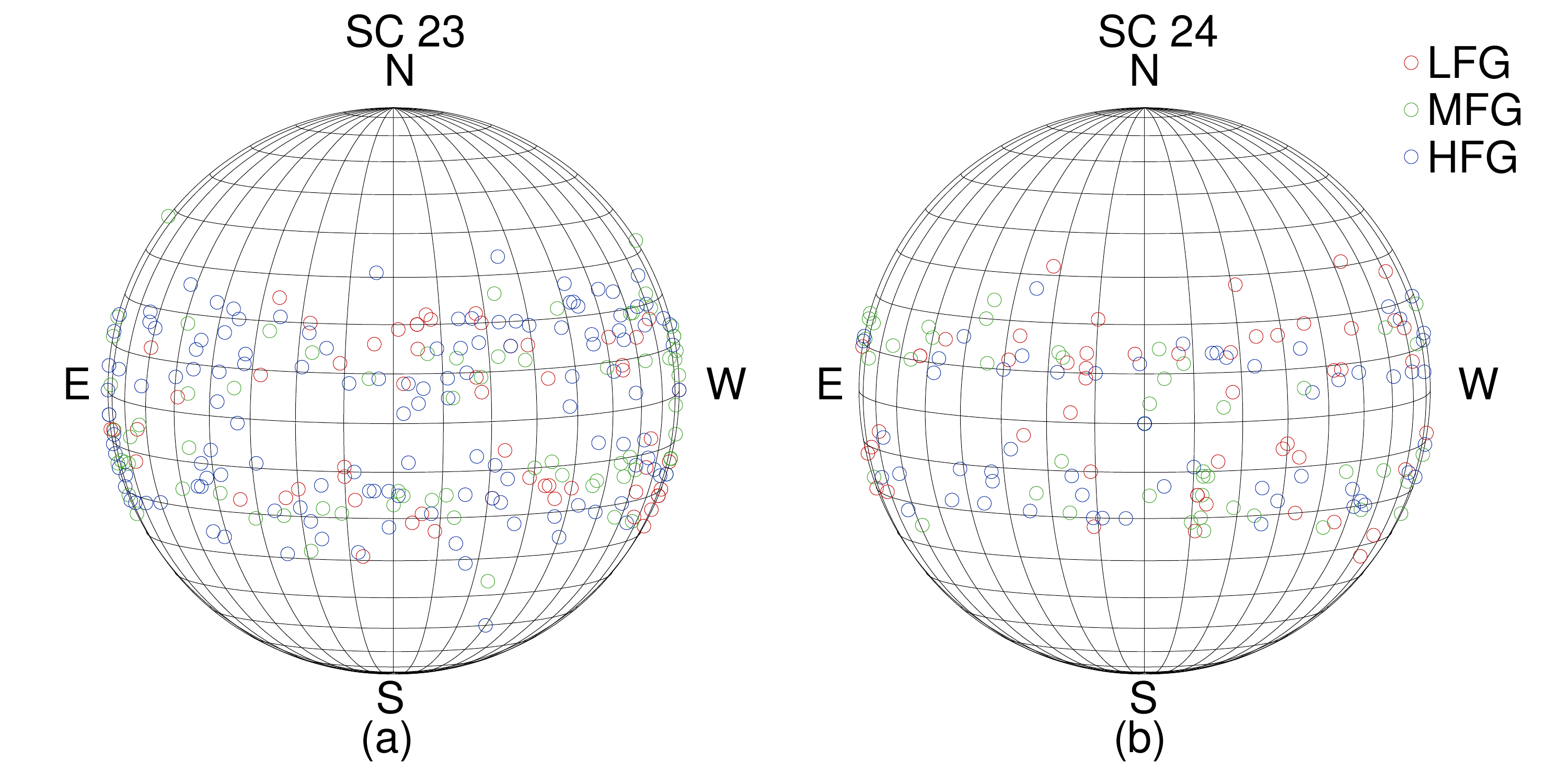}
\caption{Solar source location of the flares associated with the DH type II producing CMEs in heliographic coordinates. The grid lines on the solar disk are at every 10$^\circ $. Panels (a) and (b) represent events associated with Solar Cycles 23 and 24, respectively. The red, green, and blue symbols represents the LFG, MFG, and HFG events, respectively.}
\label{fig:location_CME}
\end{figure}

The heliographic location of DH type II bursts is shown in Figure~\ref{fig:location_CME}, separately for Solar Cycle 23 (Figure~\ref{fig:location_CME}a) and 24 (Figure~\ref{fig:location_CME}b). The solar source locations of the eruption associated with the type II bursts are taken from the Wind/WAVES type II burst catalogue \citep[for details see][]{2019SunGe..14..111G}. The events for which the exact source location is known have been considered. For Solar Cycle 23, source locations have been identified for 59 (out of 81), 73 (out of 96), and 128 (out of 158) events of LFG, MFG, and HFG categories, respectively; while for cycle 24, heliographic locations are available for 52 (out of 57), 50 (out of 54), and 61 (out of 68) events for LFG, MFG, and HFG categories, respectively.

\begin{figure}
\includegraphics[width=1.0\textwidth]{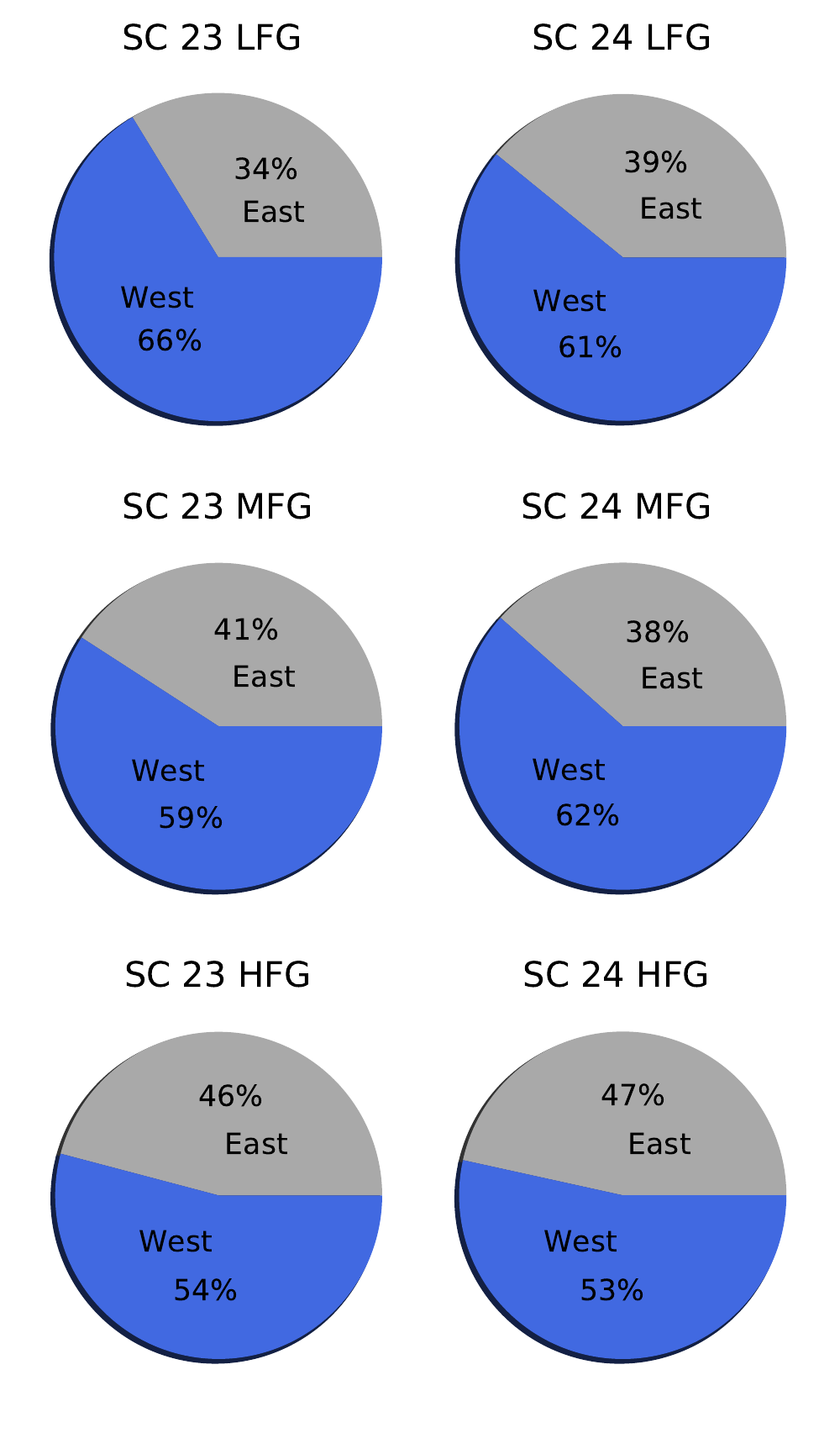}
\caption{Pie chart showing a comparison between the events occurring in the eastern (grey) and western (blue) hemispheres of the Sun for Solar Cycles 23 and 24. The events under LFG, MFG, and HFG categories are shown separately.}
\label{fig:pi_east_west}
\end{figure}

We find that there is no preferred heliographic location for the origin of LFG, MFG, and HFG events; the sources are homogeneously located for the three categories. In Figure~\ref{fig:pi_east_west}, we show the distribution of the source location of type II burst with respect to eastern and western hemispheres.
We speculate a possibility that the sources of DH type II bursts for all the three groups are predominantly located in the western hemisphere. Therefore, to evaluate the statistical significance for dominance of western dominance, we employ the binomial distribution test. Let us consider a distribution of $n$ events in 2 classes. The binomial formula gives the actual probability $P(k)$ of obtaining any particular distribution for the $k $ events in class 1 and $(n-k)$ events in class 2 \citep[see for e.g.;][]{1990A&A...229..540V, 2005A&A...431..359J}, such that
\begin{equation} \label{eq:p(k)} 
P(k)=\frac{n!}{k!(n-k)!} P^k (1-P)^{n-k}, 
\end{equation}
and the probability to get more than $d$ objects in class 1 is given by,
\begin{equation} \label{eq:prob}
 P(\geq d) =  \sum_{k=d}^{n} P(k)  .
\end{equation} 
In general, $ P(\geq d)$ $ >$ 10\% implies that the observed dominance of a particular hemisphere is statistically insignificant (i.e. occurrence of DH type II events does not have any east-west hemispheric preference); when 5\% $< $ $P(\geq d)$ $ < $ 10\%, it is marginally significant; and when $ P(\geq d)$ $<$ 5\%, we have a statistically significant result. In Table~\ref{table:e_w_binomial}, we provide the binomial probability for the observed dominance of the western hemisphere for each frequency group. We find that the dominance of western hemisphere is highly significant for only LFG and MFG groups  for Solar Cycle 23 while the dominance is marginally significant for these two groups in Solar Cycle 24. For the events of HFG group of both the solar cycles, the western dominance is statistically insignificant. In general, we believe that it is hard to comment on observed E--W asymmetry on the basis of limited availability of the data.  

\begin{table}
	\caption{The distribution of the source location of type II bursts with respect to eastern (E) and western (W) hemispheres. The counts are given for the three frequency groups (i.e. LFG, MFG, and HFG) for Solar Cycles 23 and 24, separately. In order to evaluate statistical significance of the observed E--W distribution, we provide binomial probability (Prob.) and corresponding dominant hemisphere (DH). The notations `H' and `M' mentioned in the column for `DH' denote highly and marginally significant result, respectively while dash (--) represents statistically insignificant result.}
	\begin{tabular}{cccccc}

	\hline
		Solar Cycle &Group& Hemisphere & {Number of events} & Prob. & DH\\
      \hline
      23 & LFG & E &27 & 2.4 $\times $10$^{-3}$ & W (H)\\
               && W &53 &&\\
        & MFG & E &38 & 4.82 $\times $10$^{-2}$ & W (H)\\ 
              && W& 55 &&\\
        & HFG &  E& 67 & 0.1813 &--\\
         && W &79 &&\\
      24 & LFG & E & 18 & 0.091 & W (M)\\  
             && W &28 &  & \\
          & MFG & E & 15 & 0.099 & W (M)\\
          && W & 24 & &\\
          & HFG & E &27 & 0.347  &--\\
          && W &31 &&\\

        %  \hline  

          \hline

\end{tabular}
\label{table:e_w_binomial}
\end{table}

\subsection{CMEs Associated with the DH Type II Radio Bursts}
\begin{figure}
\includegraphics[width=1.0\textwidth]{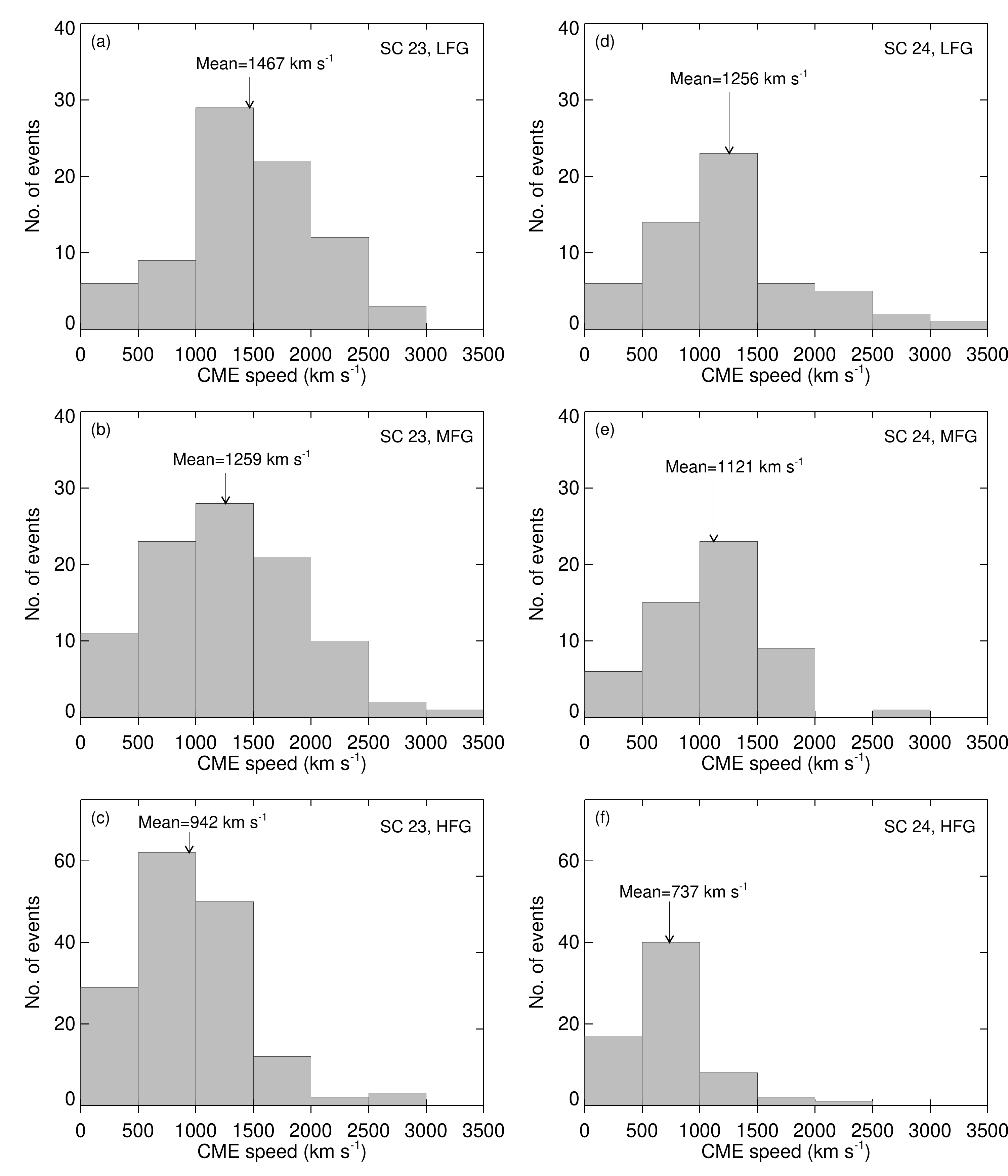}
\caption{Histogram showing the CME speed (within LASCO field of view) distribution for events of LFG (panels a and d), MFG (panels b and e), and HFG (panels c and f) categories during Solar Cycles 23 and 24.}
\label{fig:hist_CME_speed}
\end{figure}

In Figure~\ref{fig:hist_CME_speed}, we present distribution of CME speeds associated with DH type II radio bursts during Solar Cycles 23 (left panels) and 24 (right panels) for LFG, MFG, and HFG categories. The histograms clearly reveal a larger difference between mean CME speed for LFG and HFG categories. For example, in the case of Solar Cycle 23, the mean speed difference for LFG and HFG events comes out to be 525 km~s$^{-1}$ (\textit{cf.} panels a and c). Further, for both the cycles, the mean speeds for MFG events lies in-between the LFG and HFG classes. Another interesting aspect is higher mean speeds for events of Solar Cycle 23 in comparison to the next cycle for all the three frequency groups. In order to investigate whether the CME speeds for both the solar cycles belong to the same distribution or not (i.e., if the difference in mean speeds of CMEs for Solar Cycles 23 and 24 is statistically significant), we perform the two sample Kolmogorv-Smirnov test (K-S test). K-S statistic is defined as the largest absolute difference between the two cumulative distribution functions as a measure of disagreement \citep{1992nrfa.book.....P}. Suppose, the first sample has size $n_1 $ with an observed cumulative distribution function $F(x)$ and the second sample has size $n_2 $ with an observed cumulative distribution function $G(x)$, then K-S test statistic is defined as, 
\begin{equation} \label{eq:k-s_test}
 D = (\frac{n_1 n_2}{n_1 + n_2})^{1/2} |F(x)- G(x)|.
\end{equation}
 In Table~\ref{table:K-S_prob}, we present the results of K-S test (i.e. D value and probability) on the CME speeds for Solar Cycles 23 and 24. The small value of probability suggests that the null hypothesis can be rejected, i.e. the two samples are drawn from two different distributions. The two sample K-S test suggests that the difference in CME speeds for Solar Cycle 23 and 24 is statistically significant.

\begin{figure}
	\includegraphics[width=\textwidth]{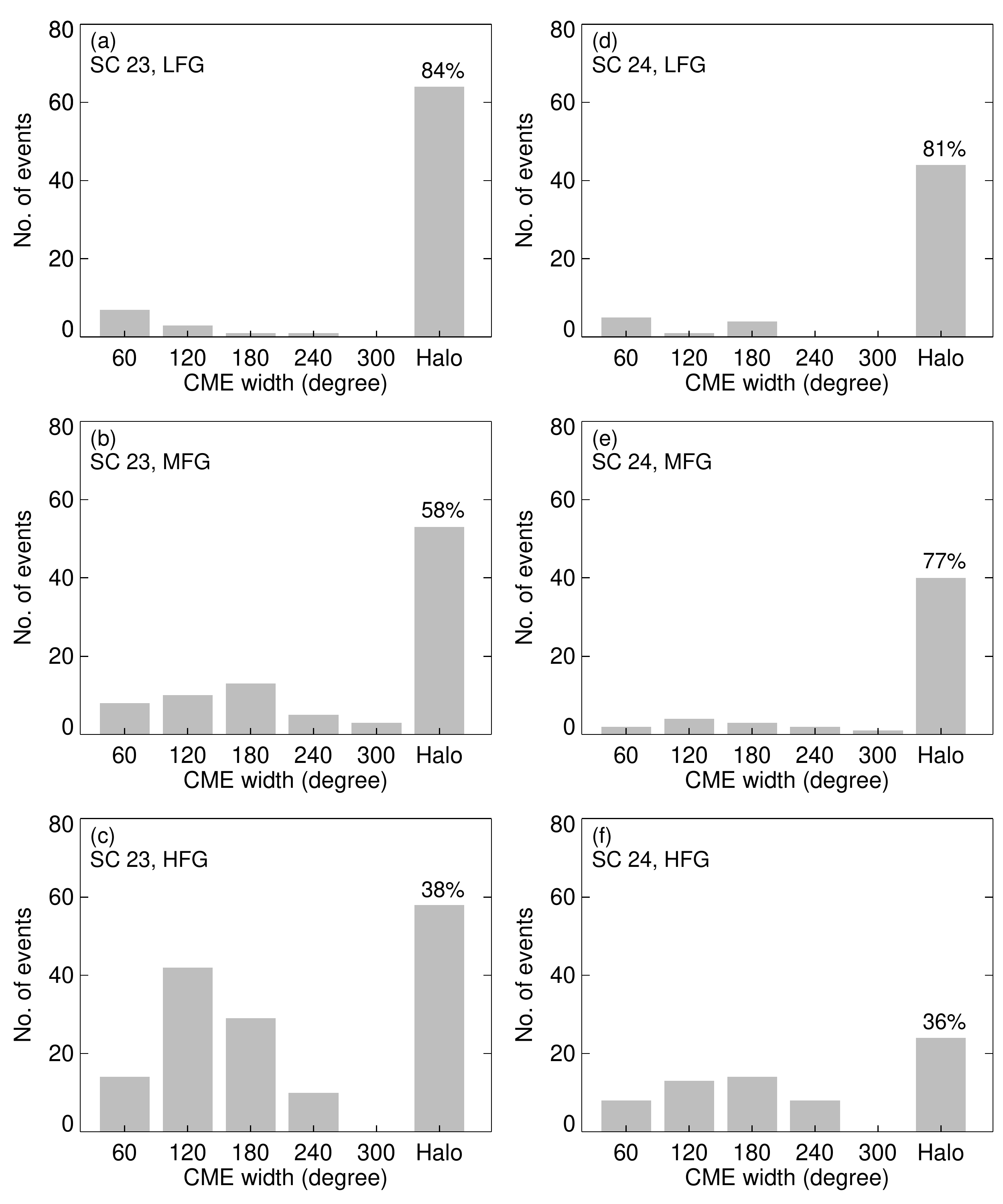}
	\caption{Histogram showing angular width of CMEs associated with DH-type II radio bursts for LFG (panels a and d), MFG (panels b and e), and HFG (panels b and d) categories Solar Cycles 23 and 24.}
\label{fig:CME_angular_width}
\end{figure}

We analyse the distribution of CME widths for Solar Cycles 23 and 24, again for the three categories of DH type II radio bursts in Figure~\ref{fig:CME_angular_width}. We find that the fraction of halo CME is significant for all the groups. Note that the fraction of halo CMEs with respect to the total events of a given category is annotated above the bar. Notably, the fraction of halo CMEs, considering all the groups together, for Solar Cycle 24 is higher than that of the Cycle 23. It is also worth mentioning that the events under HFG category, in comparison to other groups, contains largest number of partial halo and non-halo cases (see Figure~\ref{fig:CME_angular_width}c and f). In particular, the fraction of halo CMEs (36\%) is significantly reduced for HFG category of Solar Cycle 24 (Figure~\ref{fig:CME_angular_width}f).

\begin{figure}
\includegraphics[width=1.0\textwidth]{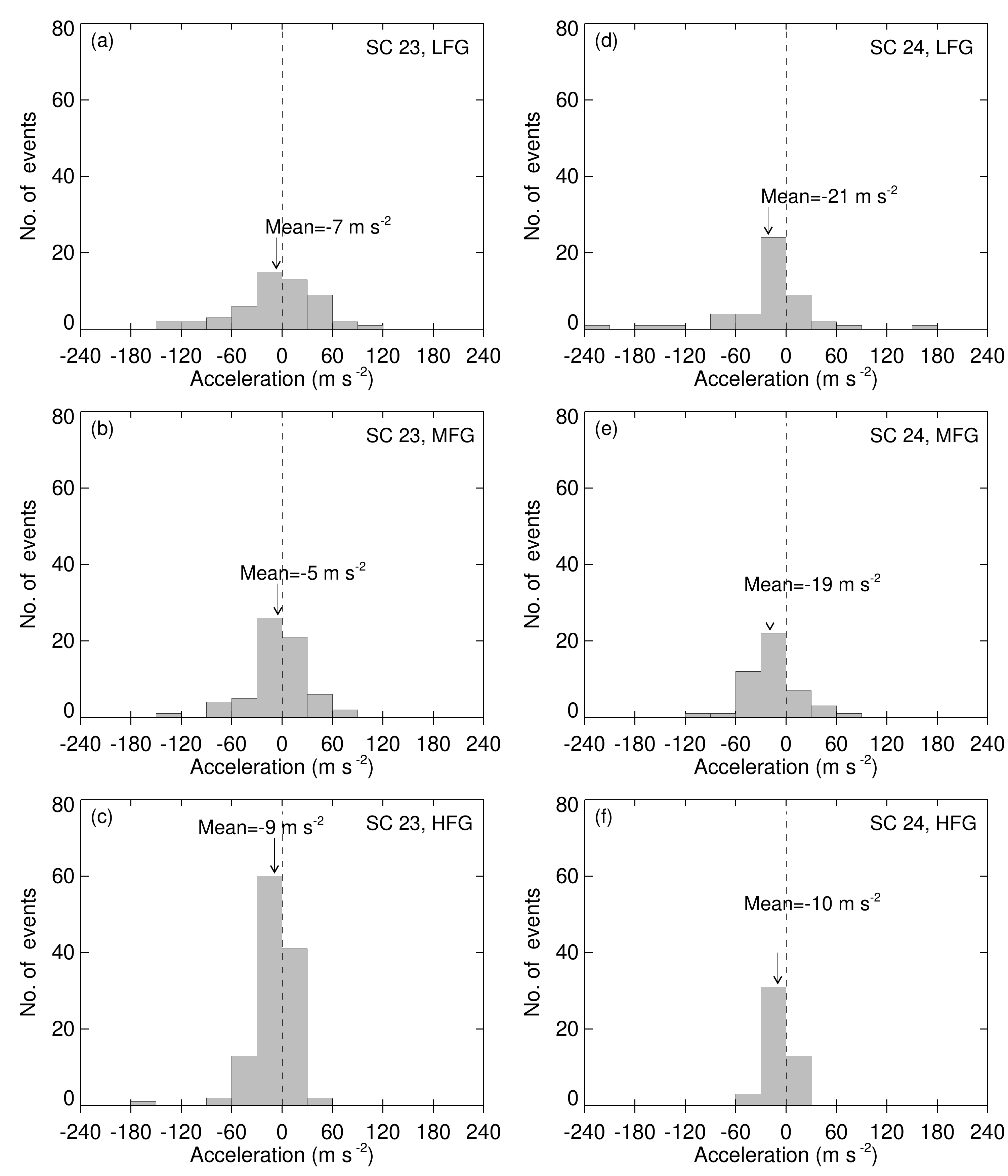}
\caption{Histogram showing the acceleration of CMEs for events of LFG (panels a and d), MFG (panels b and e), and HFG (panels c and f) categories during Solar Cycles 23 and 24. The dashed line in each panel corresponds to the zero  value.}
\label{fig:hist_CME_acc}
\end{figure}

The histograms showing the acceleration of CMEs that produce DH type II radio bursts of different categories are presented in Figure~\ref{fig:hist_CME_acc}. In most of the cases, we find that majority of DH type II associated CMEs show a mild deceleration (range of 5--10 m s$^{-2}$). Further, it is noteworthy that the LFG and MFG events of Solar Cycle 24 present higher value of mean deceleration ($\approx$20 m~s$^{-2}$; Figure~\ref{fig:hist_CME_acc}d and e).

\begin{figure}
\includegraphics[width=1.0\linewidth]{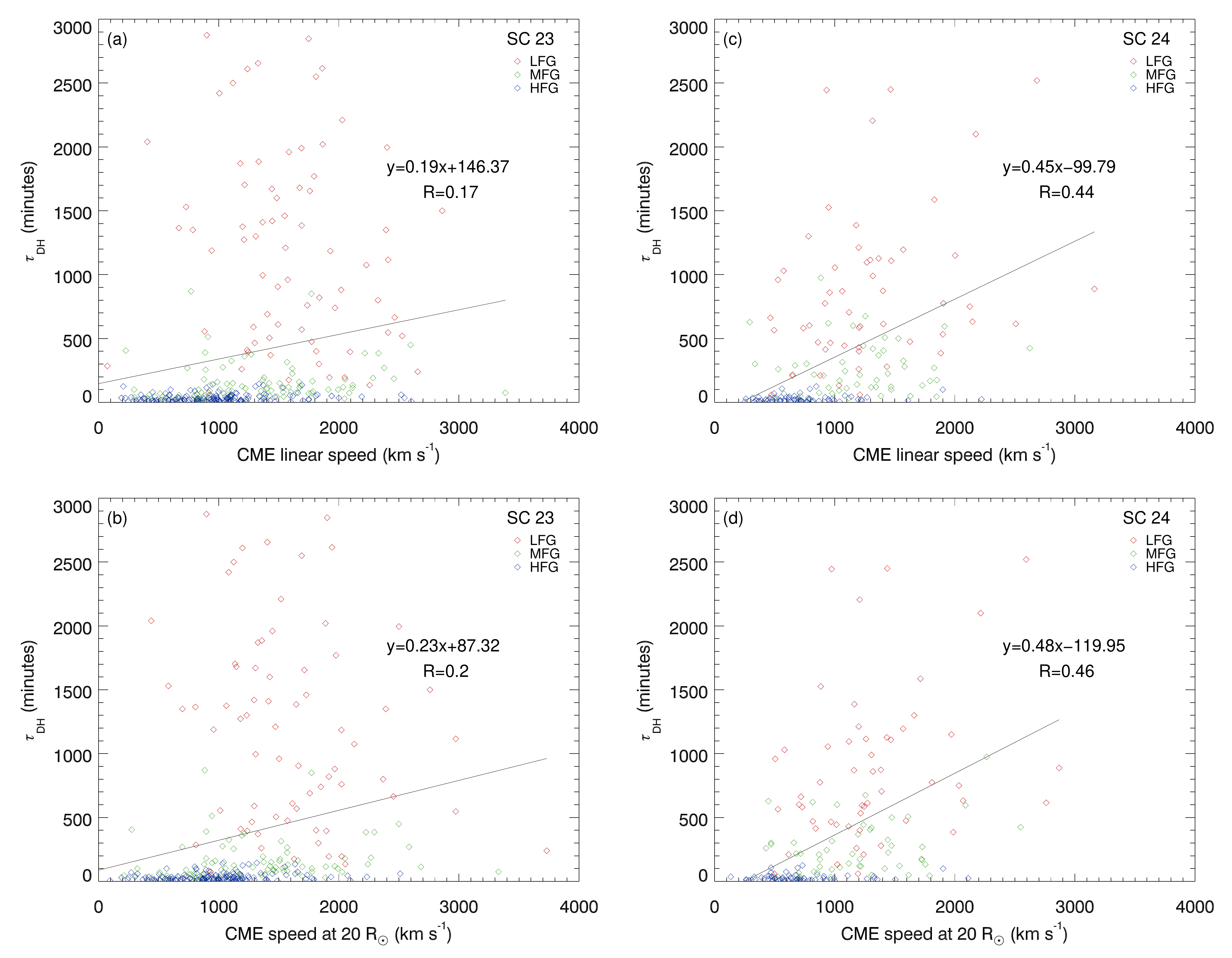}
\caption{Correlation plots for CME speeds and duration of the DH type II events 
($\tau_{\rm DH}$). We have considered CME linear speed (panel a and c) as well as CME speed at 20 R$_\odot$ (panel b and d) for Solar Cycles 23 and 24, separately. $n$ denotes number of data points in each plot. $R$ represents Pearson coefficient for the linear regression line which is also annotated in the panels. The critical values for Pearson correlation coefficients for each plot at significance level, $\alpha$=0.05, are 0.11 (panel a), 0.112 (panel b), 0.15 (panel c), and 0.152 (panel d).} 
\label{fig:CME_dur-speed}
\end{figure}

In order to explore the relation between CME speed and duration of DH type II radio bursts, we provide correlation plots between the two parameters in Figure~\ref{fig:CME_dur-speed}. For the correlation analysis, we consider two kinds of CME speeds: linear speed and speed at 20 R$_\odot$. To check the validity of the correlations, we also mention Pearson correlation coefficients obtained for the fitting of each linear regression line along with the corresponding critical values for the Pearson correlation in Figure~\ref{fig:CME_dur-speed} at significance level ($\alpha$) of 0.05. The higher values of the correlation coefficients in comparison to corresponding critical values imply that the trends shown by the linear regression line between CME speed and $\tau_{\rm DH}$ are acceptable. Importantly, the analysis reveals that the correlation between CME speed and $\tau_{\rm DH}$ is significantly improved for Solar Cycle 24 over the previous cycle. Further, we find that CME speed at 20 R$_\odot$ correlates better with event duration. 

\begin{table}
\caption{The results of two sample K-S test assessing the observed difference between mean CME speeds for Solar Cycles 23 and 24. We consider CME speeds for LFG, MFG, and HFG categories, separately. $N_1 $ and $N_2$ represent the number of events for Solar Cycles 23 and 24, respectively. The K-S test statistic and probability are denoted by D and Prob., respectively. For reference, we also provide average CME speeds.}
\begin{tabular}{ccccccc}
\hline
&  \multicolumn{2}{c} {Mean CME speed (km s$^{-1}$)} & & \\
\cline{2-3}
Solar Cycle & SC 23 & SC 24 & $N_1$ & $N_2$& D&Prob.\\
\hline
LFG & 1467 & 1256 & 70 & 54& 0.27&0.012\\
MFG & 1259 & 1121 &93 & 51&0.168 & 0.252\\
HFG &942 & 737 &152 &67 &0.34 & 2.05$\times 10^{-5}$\\
Total & 1222  & 1038  & 320 &172 & 0.158  &0.006\\
\hline
\end{tabular}
\label{table:K-S_prob}
\end{table}

\subsection{Flares Associated with DH Type II Radio Bursts}

\begin{figure}
\includegraphics[width=1.0\linewidth]{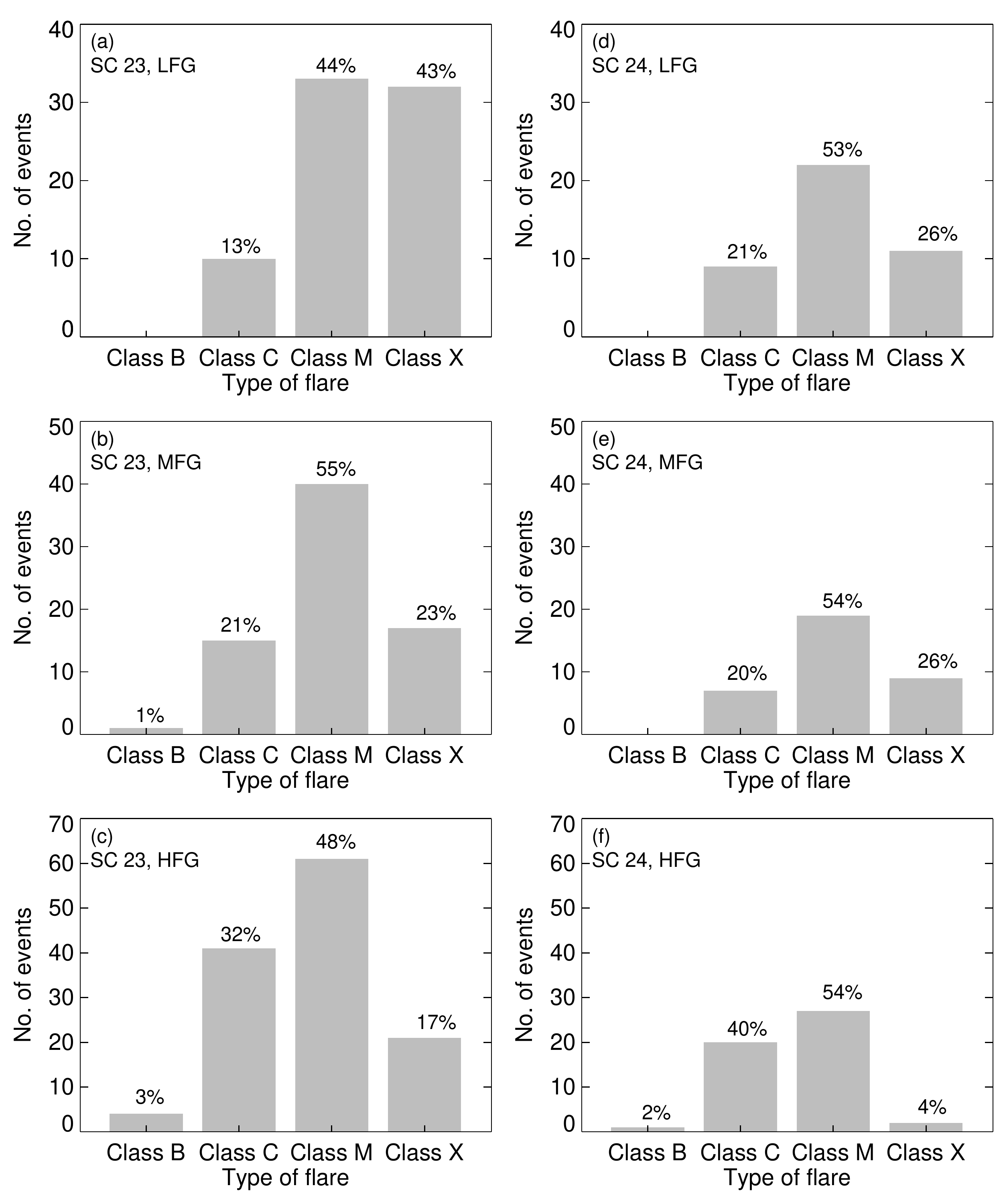}
\caption{Histograms showing the distribution of flares of GOES classes (i.e. B, C, M, and X) associated with DH-type II radio bursts. Panels (a), (b), and (c) represent LFG, MFG, and HFG categories for Solar Cycle 23, respectively. Panels (d), (e), and (f) represent LFG, MFG, and HFG categories for Solar Cycle 24, respectively.}
\label{fig:hist_flare_CMX}
\end{figure}

In Table~\ref{table:flare_as}, we provide statistics of solar flares in context of association between flare and DH type II burst. We find that for Solar Cycles 23 and 24, the clear association could be established for 82\% and 71\% cases, respectively. The reasons behind not being able to relate a DH type II burst with flare are following: the CME occurs without a flare (see column 5 of Table~\ref{table:flare_as}) or the source region of CME is located just behind the limb (see column 6 of Table~\ref{table:flare_as}) and CME is originating at the far-side (see column 7 of Table~\ref{table:flare_as}). In Figure~\ref{fig:hist_flare_CMX}, we present the histogram showing the association of flares of different GOES classes (\textit{viz} B, C, M, and X) that occur in conjunction with DH type II radio bursts. We find highest number of events are associated with the M class flares ($\approx $45--55\%) for all the categories across the two solar cycles. We further find for LFG category of Solar Cycle 23, a significant number of X class flares ($\approx $43\%) are associated with type II bursts. 

\begin{figure}
\includegraphics[width=1.0\linewidth]{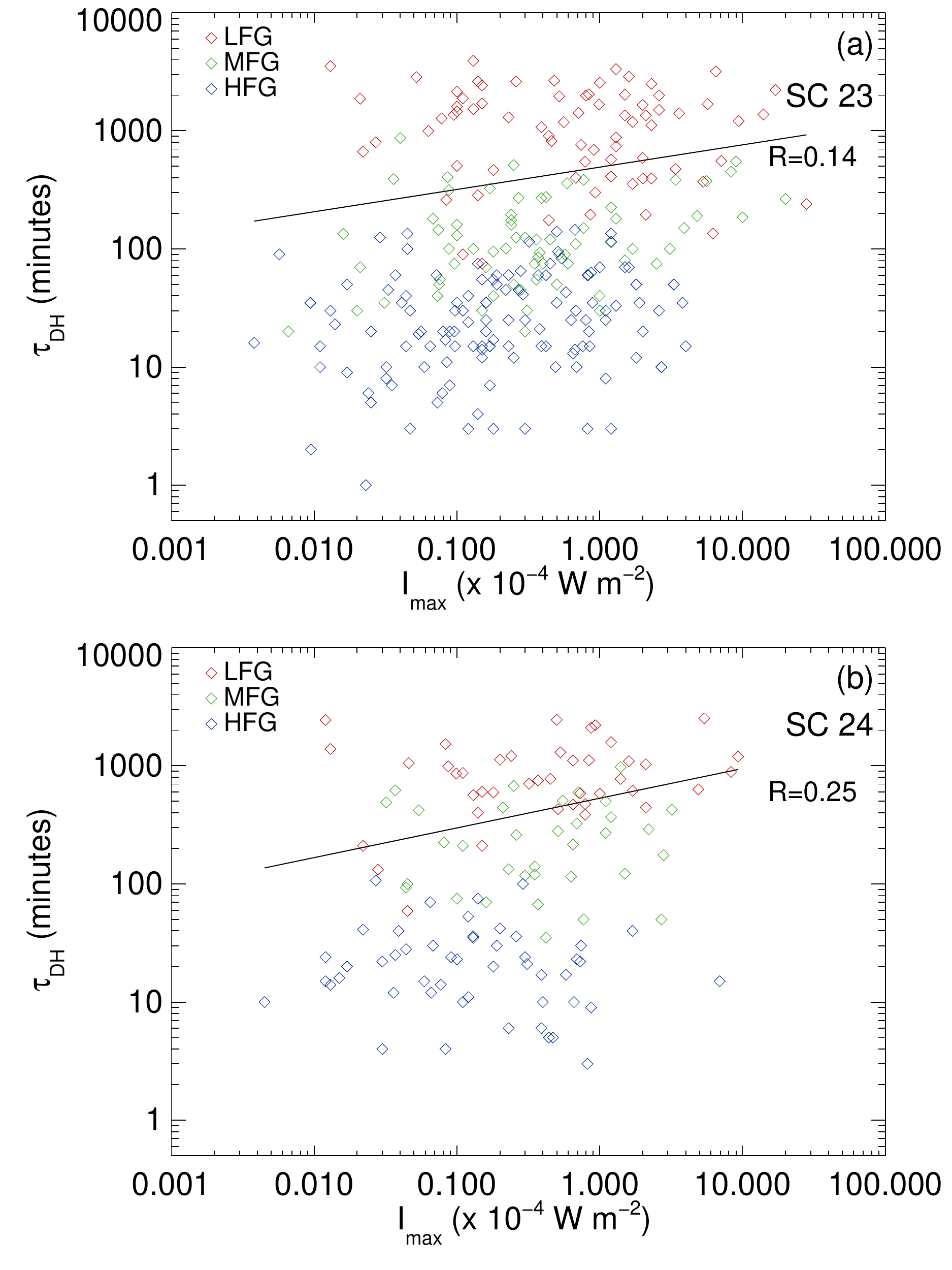}
\caption{Correlation between peak X-ray intensity ($I_{\rm max}$, in units of W m$^{-2} $) during the flare with the duration of DH type II events ($\tau_{\rm DH}$) for Solar Cycle 23 (panel a) and 24 (panel b). The red, green, and blue symbols represents the LFG, MFG, and HFG events, respectively. $n$ denotes number of data points in each plot. $R$ in each plot indicates the Pearson correlation coefficient for linear regression line. The critical values for Pearson coefficient at significance level, $\alpha$=0.05, are 0.118 (panel a) and 0.174 (panel b).} %(${\mbox{\scriptsize \tau}}_{\rm DH}$)
\label{fig:flare_flux_dur}
\end{figure}

In order to quantitatively understand whether the strength of a solar flare is related to the duration of corresponding DH type II radio bursts, we study correlation between peak X-ray intensity ($I_{\rm max}$) during the flare and duration of type II burst ($\tau_{\rm DH}$). Here, $I_{\rm max}$ denote the peak X-ray intensity during the flare in GOES 1--8 \AA~channel (in units of W m$^{-2}$). The plots in Figure~\ref{fig:flare_flux_dur} readily shows a positive yet weak correlation between the two parameters. In Figure~\ref{fig:flare_flux_dur}, we also provide the Pearson correlation coefficients obtained for the fitting of each linear regression line along with the corresponding critical values for Pearson correlation. 

The higher values of the correlation coefficients in comparison to corresponding critical values imply that the trends shown by the linear regression line between $I_{\rm max}$ and $\tau_{\rm DH}$ are acceptable. 
\begin{table}
	\caption{DH-type II bursts and flare association for all the frequency groups for Solar Cycles 23 and 24.}
	\begin{tabular}{ccccccc}

	\hline
		Solar Cycle & 	\multicolumn{6}{c} {Number of events} \\
    \cline{3-7}
		            &Group&Total & Clear type II \&  & Without  & Behind & Farside\\
		            & & &flare association &flare & the limb & \\
     \hline
        23 &LFG &81 &  75  &-- &3  &3      \\
           &MFG &96 & 73   & 1  & 18  &  4        \\
          &HFG &158 &127  & 7  &11  & 13 \\ 
          &Total & 335 & 275 (82\%)\\
         \hline
       24 & LFG &57 &  42  &-- &  2  & 13 \\
          & MFG & 54 &  35  &1 &3   &15 \\
          & HFG & 68 & 50  & 3  &  7  & 8 \\
          &Total & 179 & 127 (71\%) \\
     \hline 
	\end{tabular}
	\label{table:flare_as}
\end{table}

\subsection{CME Location at the End Frequencies of DH Type II Burst}
\label{sec:CME_loc_fe}
To analyse the locations of CMEs in the corona and interplanetary medium with respect to corresponding type II burst observations, we rely on coronal density models \citep{2007SoPh..244..167P}. Imperatively, to account for any genuine variation in coronal densities, we apply density multipliers to the models in order to aptly accommodate for drastic and transient density variations in the medium. As depicted in Figure~\ref{fig:labnalc} (see also Section~\ref{sec:origin-classification}), the maximum height for C3 field of view i.e. 30 R$_\odot$ roughly corresponds to the plasma frequency of 200 kHz. Therefore, to explore the CME height versus type II burst structures beyond LASCO FOV, a suitable CME propagation model is required. In our study, for such cases, we rely on Drag Based Model (DBM) of \cite{2013SoPh..285..295V}. This model assumes the dynamics of ICMEs is dominated by the MHD ``aerodynamic" drag. The model uses the acceleration form: 
\begin{equation}\label{eq:dbm_acc}
 a = -\gamma (v-\omega) |v-\omega|, 
\end{equation}
where $\gamma $ is the drag parameter, $v$ is the CME speed at 20 R$_\odot $ and $\omega $ is the solar wind speed. Solving this equation of motion numerically provides the heliocentric distance for ICME leading edge ($r$). The analytical solutions for the acceleration Equation~\ref{eq:dbm_acc}, with the approximation $\gamma (r)$ = constant and $\omega(r)$ = constant, gives 
\begin{equation}\label{eq:icme_speed_dbm}
v(t)=\frac{v_0-\omega}{1 \pm \gamma(v_0-\omega)t} + \omega
\end{equation}
\begin{equation}\label{eq:cme_height}
r(t) = \pm \frac{1}{\gamma} ln[1 \pm \gamma (v -\omega)t] + \omega t + r_0, 
\end{equation}
where $\pm $ depends on deceleration/acceleration regime, i.e. it is plus for v $> \omega $, and minus for v $< \omega $. Here, we note that solar wind speed and the drag parameter may vary for different solar cycles. However, for the present analysis we have taken the representative values as $\gamma$ = 1$\times 10^{-7} $ km$^{-1} $ and $\omega $ = 450 km~s$^{-1}$. Here, t is the difference between the end time of DH type II burst and onset time of corresponding CME. The first appearance time of CME in LASCO field of view is regarded as the CME onset time. The reference speed for the CME ($v$) is obtained at 20 R$_\odot$ from SOHO/LASCO CME catalogue. Substituting these values in Equation~\ref{eq:cme_height}, we get the estimated CME leading edge height for LFG events. 

\begin{figure}
\includegraphics[width=1.0\linewidth]{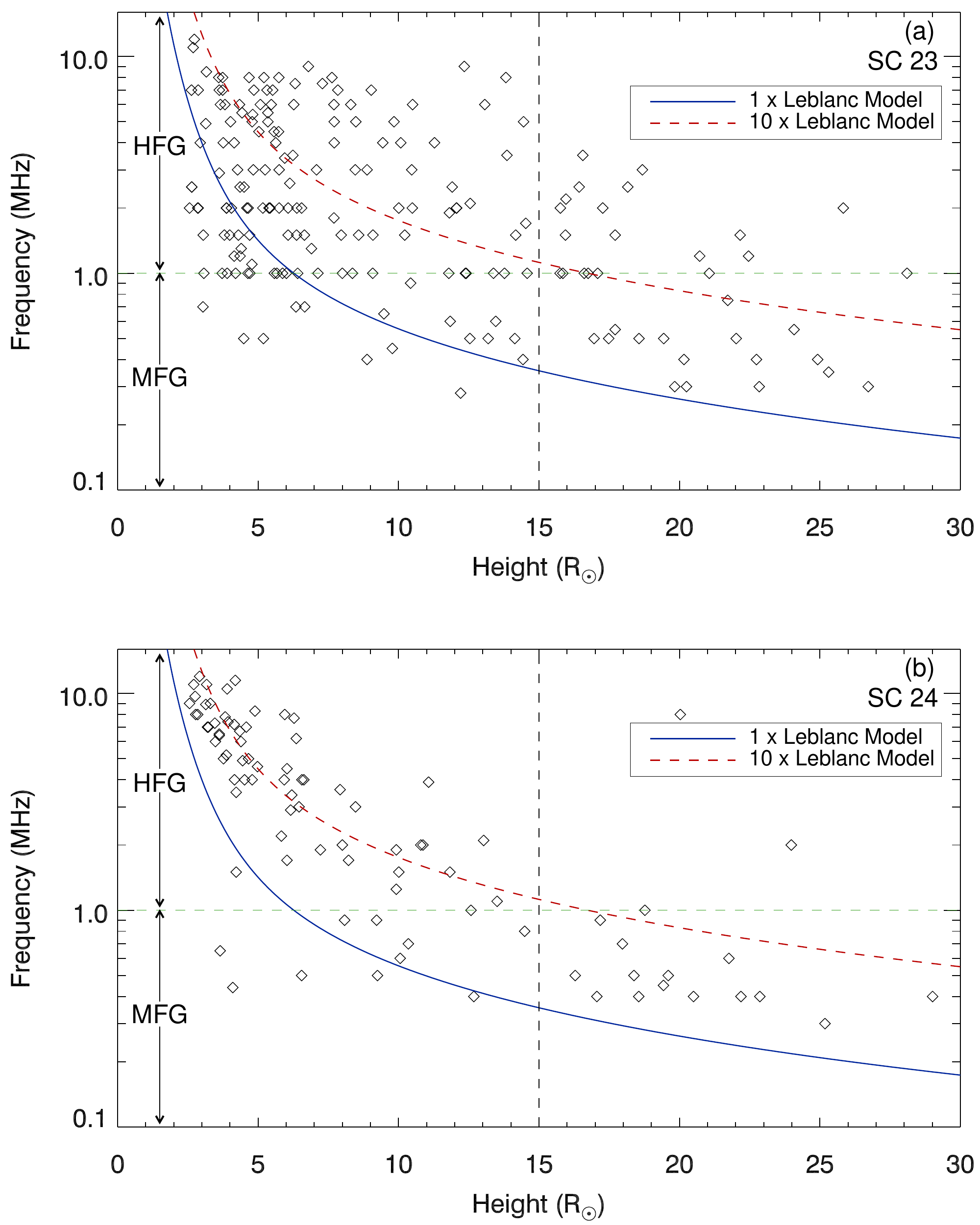}
\caption{CME height distribution at end time of DH-type II radio burst for Solar Cycle 23 (panel a) and 24 (panel b). Solid curve represents 1-fold Leblanc model and dashed curve represents 10--fold Leblanc model. The green vertical line represents the frequency of 1 MHz, which separates the HFG, and MFG categories.}
\label{fig:CME_HT_freq}
\end{figure}

In Figure~\ref{fig:CME_HT_freq}, we plot the end frequency of type II radio bursts against the height of corresponding CMEs. The plot essentially represents the projected height of CMEs, measured by LASCO, at the time when the type II structure ceases to be visible in the dynamic spectrum. In this plot, we have included all the DH type II bursts of HFG and MFG categories (200 kHz $<$ \textit{f} $\leq$ 16 MHz). The horizontal green line is drawn to separate the two categories. As noted earlier in this section, the lower frequency limit being $>$ 200 kHz for MFG and HFG categories, the direct LASCO observations concerning CME heights are feasible for the majority of events ($\approx$74\%). Figure~\ref{fig:CME_HT_freq} reveals that for a given end frequency, the CMEs exhibit a large range of heights within LASCO field of view. Further the variation in CME heights with respect to end frequencies increases for the events under MFG category compared to HFG ones. A comparison of height-frequency diagrams for cycle 23 and 24 (\textit{cf.} Figure~\ref{fig:CME_HT_freq}a and b) reveals that the HFG events of cycle 24 displays a more systematic height variations than the previous cycle and mostly confined within the profile of 10--fold Leblanc coronal density model (see Table~\ref{table:events_ht}). In general, spread of CME heights outside the curve of 10-fold Leblanc model, mostly for HFG events, indicate large variations in the coronal densities in the lower coronal heights i.e. below 6 R$_\odot$. 
The comparison between frequency-height plots for Solar Cycles 23 and 24 also indicates that for a sufficiently higher region in the corona (say 15 R$_\odot$, in Figure~\ref{fig:CME_HT_freq}a and b), the ending frequency in Cycle 24 is lower than that in Cycle 23; this implies lower coronal density for Solar Cycle 24, in comparison to the previous one.

\begin{figure}
\includegraphics[width=1.0\linewidth]{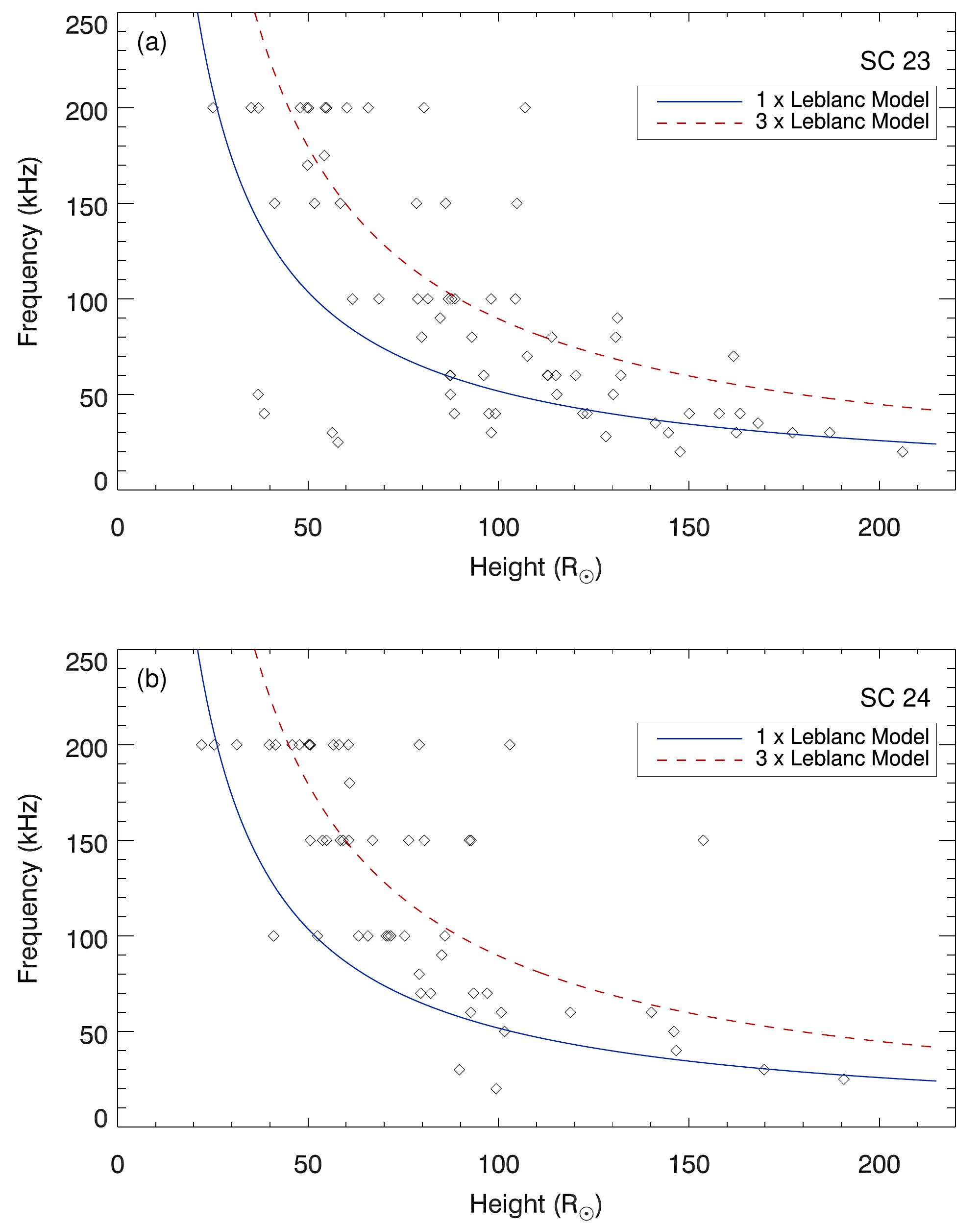}
\caption{CME height distribution at end time of DH-type II radio burst for Solar Cycle 23 (panel a) and 24 (panel b) obtained from DBM model. Solid curve represents 1--fold Leblanc model and dashed curve represent 3--fold Leblanc model.}
\label{fig:CME_DBM-HT_freq}
\end{figure}

In Figure~\ref{fig:CME_DBM-HT_freq}, we plot the end frequency against the estimated height of CMEs (based on drag based model discussed earlier in the section) for the events under LFG category. We find that compared to CMEs of HFG and MFG categories, the LFG events show much less spread in the end frequency versus height diagram (\textit{cf.} Figures~\ref{fig:CME_HT_freq} and \ref{fig:CME_DBM-HT_freq}). This feature can be well recognized by the fact that for majority of data points in Figure~\ref{fig:CME_DBM-HT_freq} are lying below the profile of 3--fold Leblanc coronal density model (see Table~\ref{table:events_DBM}) indicating smooth variations in the coronal densities in the interplanetary medium.

\begin{center}

\begin{table}
\caption{CME height at the end time of the type II burst obtained from LASCO observations for HFG and MFG events for the Solar Cycles 23 and 24.}
 
	\begin{tabular}{ccccc}

		\hline
		Solar Cycle & 	\multicolumn{4}{c} {Number of events}\\
	\cline{2-5}
&Total& Within 1-fold Leblanc & Between 1 \& 10 fold Leblanc & Outside 10 fold Leblanc\\
	& & Model & Model &Model \\
	\hline
    23 &187&33 (18\%)& 80 (42\%) & 74 (40\%)\\
	24 &88&6 (7\%)&53 (60\%) &29 (33\%)\\ 
   \hline 
	\end{tabular}
	\label{table:events_ht}
  \end{table}
 
\end{center}

\begin{center}
\begin{table}
\caption{CME height at the end time of the type II burst obtained from Drag Based model for LFG events for the Solar Cycles 23 and 24.}
\begin{tabular}{ccccc}

		\hline
		Solar Cycle & 	\multicolumn{4}{c} {Number of events}\\
	\cline{2-5}
	&Total& Within 1-fold Leblanc & Between 1 \& 3 fold Leblanc & Outside 3 fold Leblanc\\
	& & Model & Model &Model \\
	\hline
	23&73&18 (26\%) &33 (47\%)& 19 (27\%)\\
	24 &54&8 (15\%)&29 (54\%) &17 (31\%)\\ 
   \hline 
	\end{tabular}
	\label{table:events_DBM}
\end{table}
\end{center}

\section{Summary and Discussions}

In this article, we have carried out a statistical investigation of DH type II radio bursts which occurred during the time span of Solar Cycles 23 and 24. The study includes a total of 514 type II bursts out of which 65\% and 35\% events occurred during cycles 23 and 24, respectively. The study focuses on the following aspects of DH type II bursts: (1) solar cycle evolution, (2) flare-CME association, and (3) end-frequency distribution. To explore the insights of DH type II bursts in the light of above objectives, we start with classifying the observed events in three frequency domains:  Low Frequency Group (LFG; 20~kHz $\leq$~{\it f}~$\leq$ 200 kHz), Medium Frequency Group (MFG; 200 kHz $<$~{\it f}~$\leq$ 1 MHz), and High Frequency Group (HFG; 1 MHz $<$~$f$~$\le$ 16 MHz). A rational plan for this classification immediately follows from the profile presenting the variations of DH to km frequencies versus corresponding model based heliocentric distances, shown in Figure~\ref{fig:labnalc}. Evidently, a CME which manifests DH type II emission in WAVES/Wind spectrogram is expected to leave the Sun forever and subsequently enter into the IP medium \citep{2007ApJ...663.1369R}. Also, noticeable is the fact that the HFG category corresponds to the higher regions of solar corona where the topology of the magnetic field changes from closed to open field configuration \citep[e.g.,][and references within]{2013LRSP...10....5O}. Also, simple considerations based on coronal density model (e.g. 1-fold Leblanc model; Figure~\ref{fig:labnalc}) suggest that a type II burst extending below a frequency of $\approx$50~KHz represents CME--associated shock travelling beyond 0.5 AU. Hence, exploration of various frequency-wise categories of type II bursts, in their own right, is important and interesting.

The present analysis indicates that the yearly counts as well as yearly cumulative durations of DH type II show a good correspondence with solar cycle variations, manifested by sunspot numbers and areas (Figure~\ref{fig:histogram_event_dur}, \ref{fig:corr_event_ssn_sa}, \& \ref{fig:hist_DH_typeII}). The consistency in the occurrence of DH type II bursts with respect to the characteristics of solar activity cycle is further demonstrated by their heliographic source locations (Figure~\ref{fig:location_CME}) where the bursts are found to be distributed within $\pm$30$^{\circ}$ i.e. sources lie in active region belts \citep{1993SoPh..148...85H, 2005A&A...431..359J}. It is noteworthy to mention that, the DH type II parameters show better correlation with sunspot area over sunspot numbers (Figure~\ref{fig:corr_event_ssn_sa}). This interesting fact suggests that the occurrence of DH type II burst producing CME has favourable association with the magnetic complexity of the sunspot group at the source location rather than appearance of sunspot as isolated locations. Contextually, the sunspot area has been mostly viewed as a proxy for increase in the magnetic complexity of the active regions while sunspot numbers tend to show more increase with the new flux emergence away from well developed active regions \citep{2006A&A...452..647J}. 

We report some interesting features in the occurrence of DH type II bursts of LFG category during solar cycles 23 and 24. As noted in Figure~\ref{fig:pi_LFG_HFG}, the cycle 24 contains significantly higher fraction of the events of LFG category compared to cycle 23 (32$\%$ \textit{versus} 24$\%$). However, a close inspection of ending frequency distribution of LFG (Figure~\ref{fig:hist_end_frq_lfg}a and b) shows that the cycle 23 contains much larger fraction of events at frequencies $\leq$50~kHz (see dashed red lines in Figures~\ref{fig:hist_end_frq_lfg}a and b marking 50~kHz frequency). This lower frequency range (i.e. 20--50 kHz) essentially signify the shocks that are able to propagate beyond the middle of the Sun-Earth distance ($\approx $100--215 R$_\odot$; Figure~\ref{fig:labnalc}). Hence, although cycle 24 exhibits higher number of events of LFG category, it significantly lacks events that can affect near-Earth environment. These observations are consistent with the findings showing significantly lower geoeffectiveness of Solar Cycle 24 \citep{2015JGRA..120.9221G, 2021Ap&SS.366...24K}. The low counts of LFG events extending below 50~kHz during cycle 24 may be related to the lower initial speeds of CMEs during this cycle (Figure~\ref{fig:hist_CME_speed}) while the aerodynamic drag during the two cycles (i.e. 23 and 24) were found to be comparable \citep [e.g.,][]{2019SoPh..294...47S}. In this regards, we should mention that cycle 24 has slightly higher percentage of halo CMEs compared to previous cycle (65$\%$ \textit{versus} 60$\%$; see Figure~\ref{fig:CME_angular_width}). The wider CMEs are expected to experience larger aerodynamic drag in comparison to the narrower ones \citep{2010A&A...512A..43V}. 

The present study explores, in detail, the properties of eruptive phenomena at the solar source regions (i.e. flares and CMEs) associated with DH type II radio bursts. In this context, we further expand the work by \citep{2019SunGe..14..111G}. The histograms for CME speeds during both the cycles (Figure~\ref{fig:hist_CME_speed}) suggest a large difference between the mean CME speeds associated with the two groups of extreme frequency range, i.e. LFG and HFG, which implies that the CMEs associated with type II bursts displaying extended frequency emissions are launched at the Sun with much higher speeds. This result is consistent with earlier works \citep[e.g.,][]{2001JGR...10629219G, 2005JGRA..11012S07G}. Further, cycle 23 produced CMEs with significantly higher mean speeds for all the three categories (i.e. LFG, MFG, and HFG) in comparison to the next cycle. The above observation can be understood in terms of overall higher magnetic activity of cycle 23 over cycle 24 \citep{2019SoPh..294...54S}. In this context, it is worthwhile to notice that the difference in the mean CME speed for LFG and HFG categories for cycles 23 and 24 are nearly the same ($\approx$520~km~s$^{-1}$) (\textit{cf.} panels a--c and panels d--f of Figure~\ref{fig:hist_CME_speed}); this result has important implications in understanding the coupling between the magnitude of active region's magnetic flux and early evolution of CMEs. Thus, in general, the Solar Cycle 24, being a weaker magnetic cycle, produced systematically lower speed CMEs, which is also reflected in the reduced average speeds of two extreme frequency groups (i.e. LFG and HFG). The histogram for CME angular width (Figure~\ref{fig:CME_angular_width}) clearly reveals a dominance of wide CMEs in producing DH type II radio bursts for all the frequency categories and both the cycles exhibit frequent associations with halo events. In this context, the HFG group exhibits a significant deficit in halo CMEs in comparison to the other two groups. The CMEs of LFG and MFG categories are predominantly halo with the LFG events showing the highest association and significantly higher fraction of halo CMEs ($>$80$\%$).

In the context of above results, it would be worth to discuss the hierarchical relation between CME energy and wavelength range of type II radio bursts \citep{2005JGRA..11012S07G, 2006GMS...165..207G, 2010nspm.conf..108G}. \cite{2005JGRA..11012S07G}, studied  the kinematic properties of CMEs (speed, width, acceleration) by considering three population of type II bursts: (1) in the metric (m) domain with no counterparts in the DH and km domains; (2) in DH domain irrespective of counterparts in the m and km domains; (3) with counterparts in all of the wavelength domains, from m to DH to km (mkm). The analysis reveals that the average speed and halo fraction increase progressively as one goes from m to mkm type II bursts implying a progressive increase in kinetic energy \citep[see also,][]{2010nspm.conf..108G}. These works clearly imply that it is CME kinetic energy which essentially organises different populations of type II bursts.

The CMEs associated with DH type II radio bursts also found to trigger strong flares at the source regions during their activation phase. For LFG and MFG events, the predominance of M and X class flares are noteworthy (Figures~\ref{fig:hist_flare_CMX}a--b, d--e). The HFG category, on the other hand, contains significant large fraction of C-class flares besides significant reduction in X-class events (Figures~\ref{fig:hist_flare_CMX}c and f). Notably, there is a significant dominance of X-class flares in the LFG category of Solar Cycle 23 (43\%) over all the other groups (Figure~\ref{fig:hist_flare_CMX}a). 

We have also explored whether the duration of DH type II radio bursts is  related to the early kinematics of CMEs or strength (i.e. energetics) of corresponding flare. To check this hypothesis, we study correlation between the durations of DH type II burst ($\tau_{\rm DH}$) and CME initial speeds (i.e. linear as well 20 R$_\odot$ speeds; Figure~\ref{fig:CME_dur-speed}). Our analysis reveals that for Solar Cycle 24, the linear relationship between CME speed and $\tau_{\rm DH}$ clearly exists while for Cycle 23 the correlations are weak. Further, for both the cycles, the correlation slightly improves when CME speed at 20 R$_\odot$ is considered.

To study a possible link between the duration of type II bursts with flare energetics, we present correlation between the duration and peak X-ray intensity ($I_{\rm max}$) for individual events in Figure~\ref{fig:flare_flux_dur} and obtain a positive yet weak correlation between the two parameters. The above analyses imply that the CME initial speed or flare energetics are partly related with the duration of type II burst and that survival of CME associated shock is determined by multiple factors/parameters related to CMEs, flares, and state of coronal and interplanetary medium. Further, a significantly higher correlation exists between CME speed and $\tau_{\rm DH}$ for Cycle 24 over 23 is also consistent with the results of lower state of heliosphere (i.e. significant reduction in the total pressure -- magnetic $+$ plasma -- in the ambient medium into which the CMEs are ejected; see \cite{2014GeoRL..41.2673G}) and low-speed dominated background solar wind conditions during Cycle 24 \citep{2016ASPC..504...59M}.

In this study, we also explore the heliocentric location of CMEs at the time when the type II radio bursts ceased to exist and comment on the applicability of distance estimations from the models based on coronal density variations. We constrained the observed (for HFG and MFG cases) and estimated (for LFG events) CME distances in the corona and interplanetary medium by choosing suitable multipliers ($m$) to the Leblanc electron density model (Figures~\ref{fig:CME_HT_freq} and \ref{fig:CME_DBM-HT_freq}). Our results indicate that for a given end frequency, CMEs tend to show large variations in their coronal and interplanetary locations. The variation of density multipliers from 1 to 10 for HFG and MFG events implies a tremendous change in the ambient coronal densities at the near-Sun region, while for LFG cases, the multiplier range of 1--3 suggest relatively smooth density fluctuations. In particular, we note large variations in the observed height of CMEs with respect to the end frequencies for HFG group of Solar Cycle 23 (Figure~\ref{fig:CME_HT_freq}a) which points toward the complex open and closed magnetic field topologies in the corresponding height range. The comparison between frequency-height diagram for Solar Cycles 23 and 24 suggests that, for a given CME height (say 15 R$_\odot $; {\it cf.} panels a and b of Figure~\ref{fig:CME_HT_freq}) the coronal density in Cycle 24 is lower than that in Cycle 23. The result is consistent with the lower state of heliosphere in Solar Cycle 24 \citep{2014GeoRL..41.2673G}. For events occupying lower height range (for e.g. below 6 R$_\odot$; see Figure~\ref{fig:CME_HT_freq}), the possibilities of CME-streamer interactions also need to be explored on case-to-case basis \citep[e.g.,][]{2008A&A...491..873C, 2011A&A...530A..16C}. Thus, the HFG events provide us opportunity to explore the CME propagation in complex magnetic environment and within faster solar wind condition at the near-Sun region while the events of LFG category are rather intriguing for understanding the space weather phenomena.

\begin{acknowledgement}
We gratefully acknowledge the WIND/WAVES type II burst catalog which forms the basis for the present study. The LASCO CME catalog is generated and maintained at the CDAW Data Center by NASA and The Catholic University of America in cooperation with the Naval Research Laboratory. SOHO is a project of international cooperation between ESA and NASA. We further acknowledge the SOHO, STEREO, GOES, and Wind missions for their open data policy. We are grateful to the anonymous referee of the paper for providing constructive comments and suggestions that have significantly enhanced the quality and presentation of the paper.	

\noindent
{\bf Disclosure of Potential Conflict of Interest} The authors declare that they have no conflict of interest.

\end{acknowledgement}

\bibliographystyle{spr-mp-sola}

%\bibliography{sola_bibliography}  
%\newpage

\end{article}
\end{document}